\documentclass[reprint,10pt,TurnoffLineNumbers]{JASA}

\usepackage{mathtools}
\usepackage{amsmath}
\usepackage{color}
\usepackage{xcolor}
\usepackage{bm}
\usepackage{mathrsfs}
\usepackage{amssymb}
\usepackage{graphicx}
\usepackage{subfigure}
\usepackage{multirow}
\usepackage{enumerate}
\usepackage{color}

\usepackage{subfigure}

\begin{document}
%% the square bracket argument will send term to running head in
%% preprint, or running foot in reprint style.

\title[JASA/DeepFBE]{Low-latency Monaural Speech Enhancement with Deep Filter-bank Equalizer}

% ie
%\title[JASA/Sample JASA Article]{Sample JASA Article}

%% repeat as needed
\author{Chengshi Zheng}
\altaffiliation{Also at: University of Chinese Academy of Sciences, 100049, Beijing, China}
\email{cszheng@mail.ioa.ac.cn}

\author{Wenzhe Liu}
\altaffiliation{Also at: University of Chinese Academy of Sciences, 100049, Beijing, China}

\author{Andong Li}
\altaffiliation{Also at: University of Chinese Academy of Sciences, 100049, Beijing, China}

\author{Yuxuan Ke}
%\altaffiliation{Also at: University of Chinese Academy of Sciences, 100049, Beijing, China}

\author{Xiaodong Li}
\altaffiliation{Also at: University of Chinese Academy of Sciences, 100049, Beijing, China}

\affiliation{Key Laboratory of Noise and Vibration Research, Institute of Acoustics, Chinese Academy of Sciences,
	100190, Beijing, China}

% ie
%\author{Author One}
%\author{Author Two}
%\author{Author Three}

%\affiliation{}

% ie
%\affiliation{Department1,  University1, City, State ZipCode, Country}

%\altaffiliation{}

% may be added after \author{}, ie
% \altaffiliation{Also at: Department1,  University1, City, State ZipCode, Country.}

%% for corresponding author
%\email{}

%% For preprint only,
%  optional, if you want want this message to appear in upper right corner of title page
% \preprint{}

%ie
%\preprint{Author, JASA}		
\preprint{Chengshi Zheng, JASA}	
% optional, if desired:
\date{\today} 

%\modulolinenumbers
%\preto{\nolinenumbers}

\begin{abstract}
It is highly desirable that speech enhancement algorithms can achieve \textcolor{black}{good} performance while keeping low latency for many \textcolor{black}{applications}, such as digital hearing aids, acoustically transparent hearing devices, and public address systems. To improve the performance of traditional low-latency speech enhancement algorithms, a deep filter-bank equalizer (FBE) framework was proposed, which integrated a deep learning-based subband noise reduction network with a deep learning-based shortened digital filter mapping network. In the first network, a deep learning model was trained with a controllable small frame shift to satisfy the low-latency demand, i.e., $\le$ 4 ms, so as to obtain (complex) subband gains, which could be regarded as an adaptive digital filter in each frame. In the second network, to reduce the latency, this adaptive digital filter was implicitly shortened by a deep learning-based framework, and was \textcolor{black}{then} applied to noisy speech to reconstruct the enhanced speech without the overlap-add method. Experimental results on \textcolor{black}{the} WSJ0-SI84 corpus indicated that the proposed deep FBE with only 4-ms latency achieved much better performance than traditional low-latency speech enhancement algorithms in terms of \textcolor{black}{the indices such as} PESQ, STOI, and the amount of noise reduction.
\end{abstract}

%% pacs numbers not used

\maketitle

%  End of title page for Preprint option --------------------------------- %

%% See preprint.tex/.pdf or reprint.tex/.pdf for many examples
\section{\label{sec:1} Introduction}
In modern digital hearing aids \citep{proakis1996digital,popelka2016hearing}, speech enhancement plays a potentially important role in noisy environments \textcolor{black}{in improving} speech intelligibility and perceptual quality. This is because the speech reception threshold (SRT) of hearing-impaired (HI) listeners is often much higher than that of \textcolor{black}{normal-hearing (NH)} listeners, \textcolor{black}{owing to reduced} temporal and spectral resolution. In the last half-century, many efforts have been made to \textcolor{black}{reduce} noise for both monaural and \textcolor{black}{bilateral} hearing aids, so as to improve speech intelligibility, listening comfort and speech quality. For \textcolor{black}{NH} individuals, speech enhancement \textcolor{black}{usually has not been found to} improve speech intelligibility, \textcolor{black}{but it can} improve the speech perceptual quality and listening comfort by removing noise components without degrading intelligibility \citep{alcantara2003evaluation,Holube1999}. Speech enhancement has already become a preprocessing step for many systems, such as audio-visual conference systems, public address systems, speech recognition systems, and hearing assistive devices.

Deep learning-based methods have become the current state-of-the-art \textcolor{black}{for} many signal processing problems like single-channel speech enhancement. In the time-frequency (T-F) domain, \textcolor{black}{typical learning targets} can be divided into two categories, namely masking-based and mapping-based. For the former, \textcolor{black}{the} ideal binary mask (IBM) \citep{Roman03speechsegregation} and ideal ratio mask (IRM) \citep{Hummersone2014} are \textcolor{black}{the} two most widely used T-F masking targets. For the latter, \textcolor{black}{the} log-power spectrum (LPS) \citep{xu2014regression} and magnitude spectrum (MS) \citep{tan2018gated} are often chosen as the mapping target. However, these learning targets only focus on modeling the magnitude \textcolor{black}{of} the clean speech and its noisy mixture without considering phase information.\textcolor{black}{The phase information is important, especially at} low SNRs \citep{paliwal2011importance}, \textcolor{black}{and} a series of complex domain-based approaches have been proposed, which aim to reconstruct the real and imaginary (RI) parts simultaneously. For example, \textcolor{black}{a complex ratio mask (CRM) was designed to directly optimize the magnitude and phase simultaneously in the complex domain \citep{williamson2015complex}, so that} the phase can be implicitly recovered. Later, a complex-valued network was introduced to estimate the complex-valued mask \textcolor{black}{by Hu {\it{et al.}} (2020)}. More recently, \textcolor{black}{Tan {\it{et al.}} (2018) proposed using} a convolutional recurrent network (CRN) to implement complex spectral mapping (CSM), where the RI parts can be estimated simultaneously. These approaches can achieve \textcolor{black}{high} performance in theory, because both the magnitude and phase of the clean speech can be \textcolor{black}{estimated}. 

\begin{figure}[t] %% h:here t:top of page
	\centering
	\includegraphics[width=8cm]{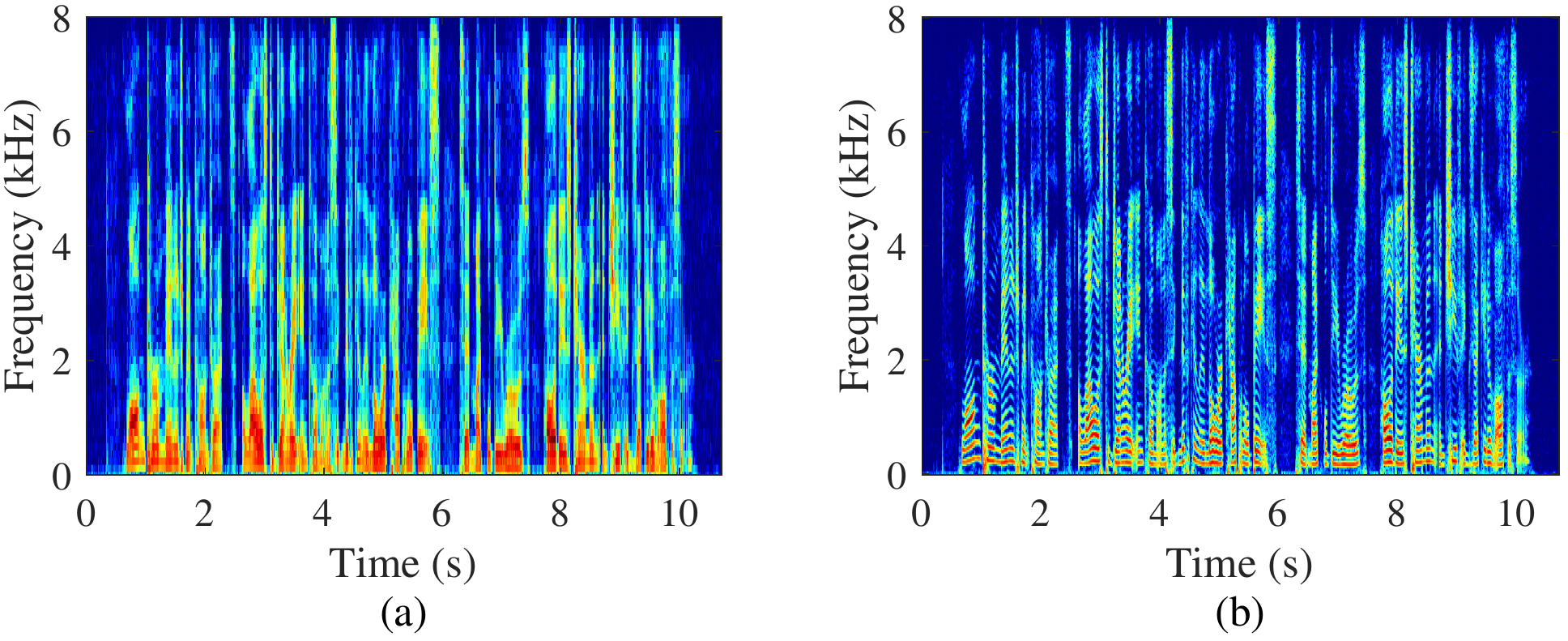}
	\caption{\label{fig:coarse_spec}\textcolor{black}{{Spectrograms of clean speech using (a) a STFT with window length 128 samples and hop length 64 samples, and (b) a filterbank with transform size 512 samples and downsampling rate 64.}}}
\end{figure}

A relatively high spectral resolution is needed in order to distinguish speech components from noise components when noise reduction algorithms are implemented in the T-F domain, where the length of the analysis/synthesis window should be large enough, e.g., 20-40 ms. This causes a relatively high system latency, since the signal delay depends on the synthesis window used for the overlap-add (OLA) reconstruction of the output. For many practical systems, such as digital hearing aids, acoustically transparent hearing devices and public address systems, low latency is required \textcolor{black}{\citep{Stone99tolerable, Stone08tolerable}}. Taking digital hearing aids as an example, the propagation time differences via \textcolor{black}{air} conduction and through the hearing aid path should be as short as possible to avoid ``coloration effects'' \citep{kates2005multichannel}. However, 
as shown in Fig.~\ref{fig:coarse_spec}, 
a reduced size of the analysis/synthesis window results in blurring and overlapping of spectral details, which \textcolor{black}{makes it harder} for deep neural networks (DNN) to learn speech spectral patterns and suppress noise components. To address this problem, \textcolor{black}{Wang {\it et al.} (2021) utilized} an asymmetric analysis-synthesis window pair to reduce the system latency for speech separation systems. However, \textcolor{black}{its latency of 8 ms is too high for most of the above mentioned systems, although it may be acceptable for digital hearing aids.} A teacher-student learning-based wave-U-Net was applied to meet the latency requirement \citep{nakaoka2021teacher}. \textcolor{black}{In this method, an offline wave-U-Net was pre-trained as the teacher model and was then used to guide the training of the student wave-U-Net model to reduce the latency with only limited performance degradation.} \textcolor{black}{It was also found that the enhancement performance worsened as the window size became smaller.}

\textcolor{black}{L{\"o}llmann and Vary (2007)} used an adaptive filter-bank equalizer (FBE) \citep{petervary2006FBE} to achieve effective speech enhancement with a controllable signal delay. The mixture signal was filtered with the adaptive FBE in the time domain to achieve much lower delay than frequency-domain filtering, while the time-domain filter coefficients were updated \textcolor{black}{with} high spectral resolution to enable the \textcolor{black}{any feasible frequency-domain speech enhancement} network to distinguish speech components from noise components. The calculation process of \textcolor{black}{the} FBE can be summarized as follows. The input signal was first decomposed into subband signals by utilizing band-pass filters with a designed prototype filter. Then the spectral gain was obtained by \textcolor{black}{traditional speech enhancement methods such as the minimum mean-square error log-spectral amplitude (MMSE-LSA) estimator, spectral subtraction, and Wiener filtering}. Finally, the output signal is obtained by filtering the mixture with the time-domain filter. Compared with spectral filtering that employs the common discrete Fourier transform (DFT) analysis-synthesis filter-bank (AS FB), it has been shown that the adaptive FBE could suppress noise in the time domain while the latency could be decreased \textcolor{black}{\citep{petervary2006FBE}}. To further decrease the signal delay, a lower degree filter was introduced \textcolor{black}{by L{\"o}llmann and Vary (2007)} to approximate the time-domain filter of the adaptive FBE, such as a moving-average (MA) filter. \textcolor{black}{The MA filter is also termed as the finite impulse response (FIR) filter commonly utilized for filtering out unwanted noise components from a time series. When the degree of the MA filter is $P$, it takes $P$ samples of the time-domain input signal and calculates the weighted sum of these samples and produces a single output sample.}

Inspired by \textcolor{black}{L{\"o}llmann and Vary (2007)}, this paper proposes a low-latency monaural speech enhancement framework with deep filter bank equalizer (DeepFBE) by combining the advantages of the deep learning-based speech enhancement method and the FBE. In the first stage, the input signal was decomposed into multiple subband signals by a set of analysis band-pass filters called the filterbank. Then the subband signals \textcolor{black}{were} downsampled by the number of subbands without any information loss, since each subband \textcolor{black}{has a limited bandwidth}. The subband-domain response of the FBE was estimated by a Noise Reduction Network (NR-Net) \textcolor{black}{using} the subband signals. Compared with traditional spectral-domain speech enhancement approaches, the NR-Net demonstrated powerful noise reduction ability by exploiting global filterbank correlations. In the second stage, a neural filter namely Filter Approximation network (FA-Net) was utilized to generate low-\textcolor{black}{order} time-domain filter coefficients to approximate the FBE subband-domain response. Then the input signal was filtered by the estimated filter via the overlap-save (OLS) synthesis method to reconstruct the time-domain enhanced output.   

The contributions of this \textcolor{black}{paper} are two-fold. First, a novel deep learning-based speech enhancement framework for low latency applications \textcolor{black}{was presented}. To the best of our knowledge, this is the first time \textcolor{black}{that} a deep learning-based FBE framework for single-channel speech enhancement \textcolor{black}{has been proposed and evaluated}. Second, we compared two neural filter design methods that approximate the FBE subband-domain response. In the time-domain signal reconstruction, the OLS method was adopted instead of the OLA method to   \textcolor{black}{reduce} the system delay.

%The rest of this paper is organized as follows. Section II presents the signal model and formulates the problem. In Section III, the proposed DeepFBE framework and its corresponding modules are \textcolor{black}{described} in detail. Section IV \textcolor{black}{describes} the experimental setup. In Section V, experimental results and analysis are given. Some conclusions are drawn in Section VI.

\section{\label{sec:2} Signal model and problem formulation}
\begin{figure*}[ht] %% h:here t:top of page
	\centering
	\includegraphics[width=16cm]{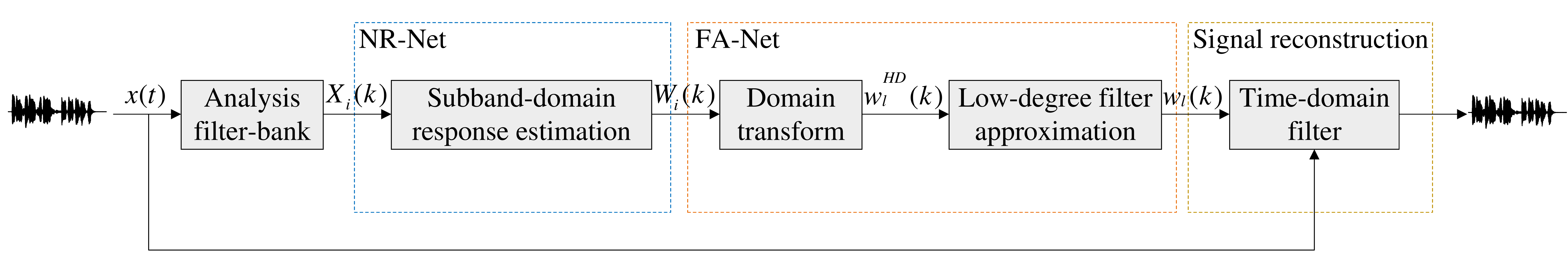}
	\caption{\label{fig:fbe}{Diagram of the adaptive filter-bank equalizer (FBE).}}
\end{figure*}
\textcolor{black}{The} single-channel noisy mixture in the time domain is modeled as:
\begin{equation}
	x(t) = s(t) + n(t),
\end{equation}
where $s(t)$ denotes the clean speech, and $n(t)$ denotes the noise with $t$ the time index. 

This paper adopts the framework for the adaptive FBE \textcolor{black}{described by L{\"o}llmann and Vary (2007)}, which can achieve aliasing-free signal reconstruction for adaptive subband filtering and has a lower delay than the corresponding DFT AS FB. 
%The analysis-synthesis process of FBE is performed as:
As shown in Fig. \ref{fig:fbe}, the $M$ subband signals $x_{i}(t)$ are generated by means of $M$ band-pass filters with impulse responses $h_{i}(l)$ of length $L+1$, \textcolor{black}{given by}
\begin{equation}
\label{equation:equation2}
x_{i}(k) = \sum_{l=0}^{L} {x(kr-l)h_{i}(l)}; i=0,1,\cdots, M-1,
\end{equation}
where $k$ is the frame index and $r$ is the downsampling rate. The impulse response of the $i$-th band-pass filter, $h_i(l)$, is a modulation of a prototype low-pass filter of length $L+1 \ge M$ with the impulse response $h(l)$, \textcolor{black}{given by}
\begin{equation}
\label{equation:equation3}
h_{i}(l) = 
\left\{
\begin{aligned}
&h(l)\phi_{i}(l),&l\in\left\{0, 1,\cdots, L\right\};i\in\left\{0, 1,\cdots, M-1\right\}\\
&0,&{\rm else}   
\end{aligned}
\right.
\end{equation}

In this work, we consider the generalized DFT (GDFT) with evenly-stacked frequency channels. Therefore, the impulse response of the prototype filter $h(l)$ and the general modulation sequence $\phi_{i}(l)$ are, respectively, given by:
\begin{equation}
h(l)=\frac{1}{M} \frac{\sin(\frac{2\pi}{M}(l-\tau))}{\frac{2\pi}{M}(l-\tau)}win(l)
\end{equation}
and
\begin{equation}
\phi_{i}(l)=\exp \left(-j\frac{2\pi}{M}i(l-\tau)\right), i=0,1,\cdots,M-1;l\in \mathbb{Z},
\end{equation}
where $win(l)$ is the Hanning window, and $\tau = L/2$ is utilized to guarantee \textcolor{black}{that} the coefficients of the FIR filter have non-zero phase. \textcolor{black}{This is important because a zero-phase FIR filter makes this system noncausal.}

The enhanced signal is synthesized with the filter-bank summation (FBS) method as:
\begin{equation}
	\label{equation:equation6}
	\widehat{s}(t) = \sum_{i=0}^{M-1}{\widehat{W}_{i}(k)x_{i}(k)},
\end{equation}
where $\widehat{W}_{i}(k)$ is the $i$-th subband response of the $k$th frame.

\textcolor{black}{From} Eqs. (\ref{equation:equation2}) and (\ref{equation:equation3}), Eq. (\ref{equation:equation6}) can be expressed as:
\begin{equation}
	\label{equation:equation7}
	\widehat{s}(t) = \sum_{l=0}^{L}{x(t-l)\widehat{w}_{l}(k)},
\end{equation}
where $\widehat{w}_{l}(k)$ is the corresponding time-domain response of the estimated subband-domain response $\widehat{W}_{i}(k)$.

\textcolor{black}{The FBE is used to produce time-domain filter coefficients that can be updated in the subband-domain. The desired subband-domain response can be estimated by} some speech enhancement algorithms, such as spectral subtraction, i.e., $\widehat{W}_{i}(k)=\mathcal{F}_{1}({x}_{i}(k))$, and the mapping between the subband-domain response and the time-domain high-\textcolor{black}{order} (HO) response is defined as $\widehat{w}^{HD}_{l}(k)=\mathcal{G}_{1}(\widehat{W}_{i}(k))$:
\begin{equation}
	\widehat{w}^{HD}_{l}(k) = \mathcal{G}_{1}(\widehat{W}_{i}(k)) = h(l)\sum_{i=0}^{M-1}{\widehat{W}_{i}(k)\phi_{i}(l)},
\end{equation}

In order to further reduce signal delay, the response can be approximated by a lower-degree filter which is denoted $\mathcal{G}_{2}(\cdot)$. The final time-domain response $\widehat{w}_{l}(k)$ can be obtained after domain transformation $\mathcal{G}_{1}(\cdot)$ and the low-\textcolor{black}{order} filter approximation operation $\mathcal{G}_{2}(\cdot)$. The function \textcolor{black}{for changing} from the subband-domain response to the low-\textcolor{black}{order} time-domain response can be defined as $\widehat{w}_{l}(k) = \mathcal{F}_{2}(\widehat{W}_{i}(k))$. 

To summarize, the implementation of the FBE consists of two stage: (1) Subband response estimation for noise reduction: in this stage, the mixture was decomposed into subband signals by the analysis filterbank with downsampling. Then the subband signals were \textcolor{black}{used} to calculate the subbband-domain response of the noise-reduction filter. (2) Filter length shortening for latency reduction: in this stage, time-domain filter coefficients were obtained by a domain transform function and a low-\textcolor{black}{order} filter approximation operation. After that, the enhanced speech was generated by filtering the input signal with the estimated time-domain filter coefficients.
\section{\label{sec:3} Proposed Two-stage Framework}
\begin{figure*}[ht] %% h:here t:top of page
	\centering
	\includegraphics[width=16cm]{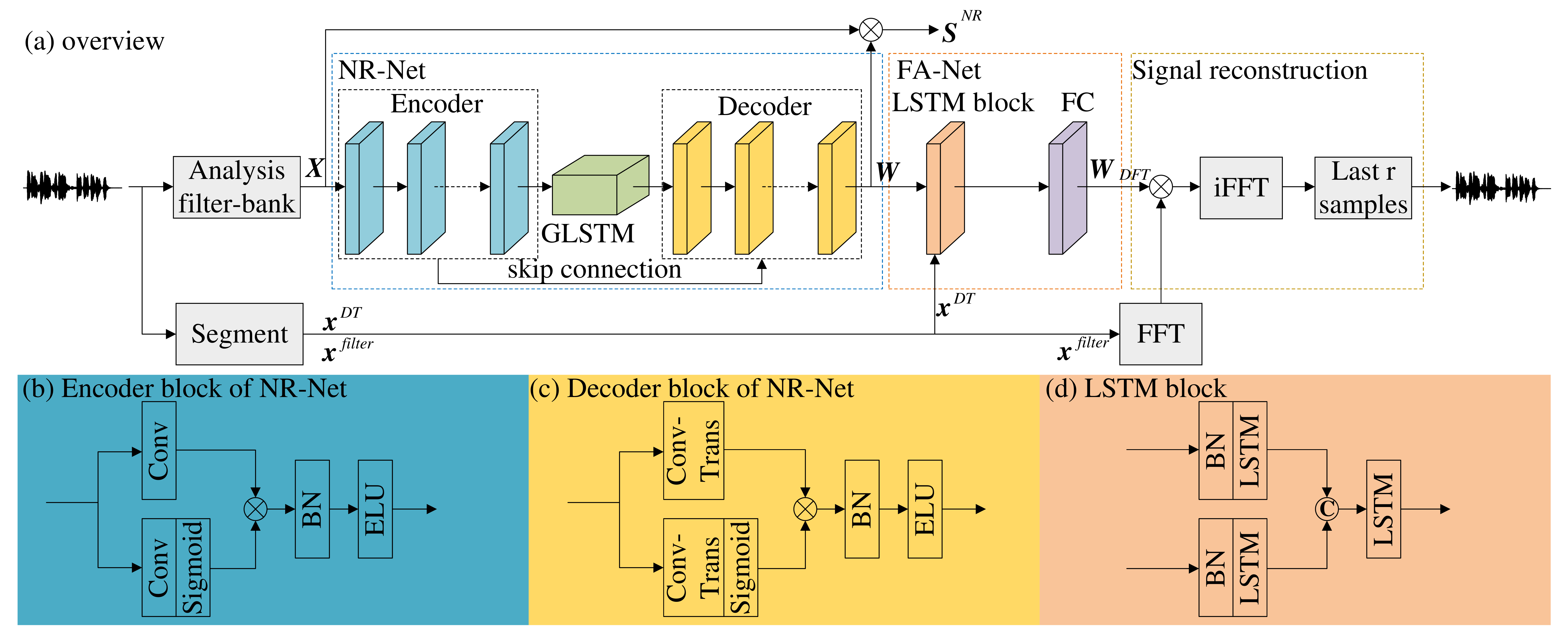}
	\caption{\label{fig:framework}{Diagram of the proposed system.}}
\end{figure*}
This section proposed a low-latency monaural speech enhancement framework with filter-bank equalizer that performed time-domain filtering with coefficients updated in the subband-domain by \textcolor{black}{a DNN}. Compared with the conventional FBE described in Section \ref{sec:2}, there are three differences: (1) A masking-based neural network was utilized to estimate a subband-domain response which achieved better performance than traditional methods especially for low SNRs and non-stationary noise scenarios. (2) The filter approximation stage was replaced by a DNN in order to use a data-dependent method to fit the high-\textcolor{black}{order} filter rather than a fixed filter design method such as the MA filter approximation. \textcolor{black}{In addition, we explored the effect of replacing the domain transformation mapping function with a network for the filter approximation} and demonstrated better performance. (3) We implemented the filtering operation in the frequency domain with the OLS method instead of the OLA method when reconstructing the time-domain signal.

Figure~{\ref{fig:framework}}(a) \textcolor{black}{is a} diagram of the proposed system, which consists of two stages, namely the Noise Reduction Network (NR-Net) and the Filter Approximation Network (FA-Net). The pipeline goes as follows. In the first stage, the full-band input signal $x(t)$ with discrete time $t$ was decomposed into $M$ subband signals $x_{i}(k)$ with $i\in\left\{0, 1,\cdots, M-1\right\}$. To reduce complexity, similar to \textcolor{black}{Vary (2006)}, this subband decomposition was achieved via a polyphase network (PPN) implementation and downsampling in the analysis filterbank. After that, the NR-Net was employed to estimate the subband filter $\widehat{\boldsymbol{W}}\in \mathbb{C}^{K \times M}$ to suppress the noise, where $K$ is the number of frames. 

To obtain time-domain filter coefficients for the full-band signal, in the second stage, the noisy signal $x(t)$ was split into frames of $M$ samples with frame shift $r$:
\begin{equation}
	\boldsymbol{x}^{DT}(k) = x[kr-M+1:kr], k \in \mathbb{Z},
\end{equation}
where $x[a:b]$ means that samples of $x(t)$ from time index $a$ to time index $b$ are included.
 
For the $k$th frame, the FA-Net received both the original noisy signal $\boldsymbol{x}^{DT}(k) \in \mathbb{R}^{M \times 1}$ and \textcolor{black}{the} estimated response $\widehat{\boldsymbol{W}}(k)$ as the input and aimed to \textcolor{black}{predict} the corresponding frequency-domain response $\widehat{\boldsymbol{W}}_{DFT}(k) \in \mathbb{C}^{D \times 1}$, where $D$ depends on the degree of the approximate \textcolor{black}{time-domain} filter. Then the mixture signal $\boldsymbol{x}^{filter}(k) = x[kr-2P+1: kr]$ was filtered by the estimated $k$th frame frequency-domain response $\widehat{\boldsymbol{W}}_{DFT}(k)$ and the last $r$ samples of the output were preserved as the enhanced output of the $k$th frame $\widehat{\boldsymbol{s}}(k) \in \mathbb{R}^{r \times 1}$. The final enhanced signal was constructed \textcolor{black}{using} the OLS method. The whole forward calculation process was formulated as:
\begin{equation}
	\widehat{\boldsymbol{W}} = \mathcal{F}_{1}( \boldsymbol{X}; \Theta_{1}),
\end{equation}
\begin{equation}
	\widehat{\boldsymbol{W}}_{DFT}(k) = \mathcal{F}_{2}( \boldsymbol{x}^{DT}(k), \widehat{\boldsymbol{W}}(k); \Theta_{2} ),
\end{equation}
\begin{equation}
	\widehat{\boldsymbol{s}}(k) = \mathcal{F}_{3}(\widehat{\boldsymbol{W}}_{DFT}(k), \boldsymbol{x}^{filter}(k)),
\end{equation}
where ${\bf X} \in {\mathbb{C}}^{K \times M}$ with ${\bf X}_{i,k} = x_i(k)$. $\mathcal{F}_{1}$, $\mathcal{F}_{2}$, and $\mathcal{F}_{3}$ denote the calculation \textcolor{black}{operations} of NR-Net, FA-Net, and signal reconstruction, respectively. $\Theta_{1}$ and $\Theta_{2}$ are the network parameters of NR-Net and FA-Net, respectively. 
\subsection{\label{subsec:3:2} Noise Reduction Network} 
\label{DNN}
A diagram of the Noise Reduction Network (NR-Net) is shown in Fig.~{\ref{fig:framework}}(b) and Fig.~{\ref{fig:framework}}(c). It is a convolutional recurrent network (CRN)\textcolor{black}{, which} has been successfully applied \textcolor{black}{in} the field of speech enhancement~{\cite{tan2019learning}}. 

The NR-Net is essentially an encoder-decoder structure with long short-term memory (LSTM) layers between the encoder and the decoder \textcolor{black}{\citep{tan2019learning}}. The input is the 257-dimensional subband complex spectrum of the input signal. The encoder includes six convolutional blocks to extract \textcolor{black}{spectro-temporal} features from the noisy spectra, and the decoder has six deconvolutional blocks to gradually \textcolor{black}{interpolate and recover the original size of the input} and estimate the complex mask. Each (de)convolutional block comprises a (De)Conv-GLU layer, batch normalization (BN) and an ELU activation layer. In order to obtain a causal system for real-time processing, we applied causal convolutions in the time dimension to the encoder and decoder layers, which were utilized to control the output at time $t$ based only on the input features from time $t$ and earlier in the previous layer, meanwhile maintain the temporal order of the input sequence. Note that causal deconvolutions can be easily applied to the decoder layers because the deconvolution is essentially a convolution operation. Within each layer, the kernel size was set to $1\times3$ except \textcolor{black}{for the first layer, where it was  $1\times5$} and the stride was $1\times2$ along the time and frequency directions to keep \textcolor{black}{all layers have the same time dimension} to meet the real-time requirement and obtain a large-range receptive field \textcolor{black}{along the} frequency axis to learn the characteristics of inter-harmonic features. \textcolor{black}{The number of channels of each convolutional block in the encoder was} set to {16, 32, 64, 64, 128, 256} from the first to the last one. The decoder was designed as a mirror version of the encoder, and \textcolor{black}{the number of channels} of the final deconvolutional layer was set to 2 without normalization and activation function to estimate the complex subband-domain mask. In this way, low-resolution feature embedding was transformed by the encoder and the decoder restored high-level features to the spectra with the original input shape. To mitigate the gradient disappearance problem, the skip connection strategy was adopted, which concatenated the output of each encoder layer and that of the corresponding decoder layer. \textcolor{black}{Between the encoder and the decoder, two LSTM layers were inserted to capture temporal sequence dependencies. To reduce the model complexity, a grouping strategy was adopted for each LSTM layer (dubbed GLSTM), where the group number was set to 4}.

\textcolor{black}{A more} detailed description of NR-Net is presented in Table~{\ref{tbl:crn}}. The input size and the output size are given in $(ChannelNum \times TimeSteps \times frequencySize)$ for both the encoder and the decoder, and $(TimeSteps \times frequencySize)$ for LSTMs. The hyper-parameters for CNNs are specified with $(KernelSize, Strides, ChannelNum)$ format.

{
\begin{table*}[t]
	\small
	\centering
	\caption{\label{tab:Architecture_of_CRN} Architecture of the NR-Net model used in this paper.}
	
	\tabcolsep22pt
	\renewcommand\arraystretch{0.8}
	\begin{tabular}{c|c|c|c}
				\hline\hline
				layer name & input size & hyper-parameters & output size\\
				\hline
				Conv\_1 & $C$ $\times$ $T$ $\times$ 257 & 1 $\times$ 5, (1, 2), 16 & 16 $\times$ $T$ $\times$ 127\\
				\hline
				Conv\_2 & 16 $\times$ $T$ $\times$ 127 & 1 $\times$ 3, (1, 2), 32 & 32 $\times$ $T$ $\times$ 63\\
				\hline
				Conv\_3 & 32 $\times$ $T$ $\times$ 63 & 1 $\times$ 3, (1, 2), 64 & 64 $\times$ $T$ $\times$ 31\\
				\hline
				Conv\_4 & 64 $\times$ $T$ $\times$ 31 & 1 $\times$ 3, (1, 2), 64 & 64 $\times$ $T$ $\times$ 15\\
				\hline
				Conv\_5 & 64 $\times$ $T$ $\times$ 15 & 1 $\times$ 3, (1, 2), 128 & 128 $\times$ $T$ $\times$ 7\\
				\hline
				Conv\_6 & 128 $\times$ $T$ $\times$ 7 & 1 $\times$ 3, (1, 2), 256 & 256 $\times$ $T$ $\times$ 3\\
				\hline
				reshape\_size\_1& 64 $\times$ $T$ $\times$ 4 & - & T $\times$ 768\\
				\hline
				GLSTM\_1 & $T$ $\times$ 768 & 768 & T $\times$ 768\\
				\hline
				GLSTM\_2 & $T$ $\times$ 768 & 768 & T $\times$ 768\\
				\hline
				reshape\_size\_2 & $T$ $\times$ 768 & - & 256 $\times$ $T$ $\times$ 3\\
				\hline
				skip\_connection\_1 & 256 $\times$ $T$ $\times$ 3 &-& 512 $\times$ $T$ $\times$ 3\\
				\hline
				DeConvGLU\_1 & 512 $\times$ $T$ $\times$ 3 & 1 $\times$ 3, (1, 2), 128 & 128 $\times$ $T$ $\times$ 7\\
				\hline
				skip\_connection\_2 & 128 $\times$ $T$ $\times$ 7 &-& 256 $\times$ $T$ $\times$ 7\\
				\hline
				DeConvGLU\_2 & 256 $\times$ $T$ $\times$ 7 & 1 $\times$ 3, (1, 2), 64 & 64 $\times$ $T$ $\times$ 15\\
				\hline
				skip\_connection\_3 & 64 $\times$ $T$ $\times$ 15 &-& 128 $\times$ $T$ $\times$ 15\\
				\hline
				DeConvGLU\_3 & 128 $\times$ $T$ $\times$ 15 & 1 $\times$ 3, (1, 2), 64 & 64 $\times$ $T$ $\times$ 31\\
				\hline
				skip\_connection\_4 & 64 $\times$ $T$ $\times$ 31 &-& 128 $\times$ $T$ $\times$ 31\\
				\hline
				DeConvGLU\_4 & 128 $\times$ $T$ $\times$ 31 & 1 $\times$ 3,(1, 2), 32 & 32 $\times$ $T$ $\times$ 63\\
				\hline
				skip\_connection\_5 & 32 $\times$ $T$ $\times$ 63 &-& 64 $\times$ $T$ $\times$ 63\\
				\hline
				DeConvGLU\_5 & 64 $\times$ $T$ $\times$ 63 & 1 $\times$ 3, (1, 2), 16 & 16 $\times$ $T$ $\times$ 127\\		
				\hline
				skip\_connection\_6 & 16 $\times$ $T$ $\times$ 127 &-& 32 $\times$ $T$ $\times$ 127\\
				\hline
				DeConvGLU\_6 & 32 $\times$ $T$ $\times$ 127 & 1 $\times$ 5, (1, 2), 2 & 2 $\times$ $T$ $\times$ 257\\		
				\hline
				\hline
			\end{tabular}
			\label{tbl:crn}
	\end{table*}
}

\subsection{\label{subsec:3:3} Filter Approximation Network}
The Filter Approximation Network (FA-Net) is shown in Fig.~{\ref{fig:framework}}~(d). \textcolor{black}{It} consists of stacked LSTM layers. The segmented noisy signal and the predicted complex subband-domain response were taken as the input, \textcolor{black}{and} passed through a batch normalization layer and an LSTM layer to generate the embedding $\mathbf{X}^{e}$ and $\mathbf{W}^{e}$, \textcolor{black}{given by}
\begin{equation}
	\mathbf{X}^{e} = \mathcal{H}_{x}(BN(\boldsymbol{x}^{DT}(k))),
\end{equation}
\begin{equation}
\mathbf{W}^{e} = \mathcal{H}_{w}(BN(\widehat{\boldsymbol{W}}(k))),
\end{equation}
where $\mathcal{H}_{x}(\cdot)$ and $\mathcal{H}_{w}(\cdot)$ are the mapping functions defined by the LSTM, and $BN(\cdot)$ is the batch normalization. Then we concatenated $\mathbf{X}^{e}$ and  $\mathbf{W}^{e}$ and sent to an LSTM layer to create an embedding \textcolor{black}{function} $\mathbf{E}^{e}$. 
\begin{equation}
\mathbf{E}^{e} = \mathcal{H}([\mathbf{X}^{e}, \mathbf{W}^{e}]),
\end{equation} 
where $\mathcal{H}(\cdot)$ is the LSTM function, and $[\cdot,\cdot]$ represents the concatenate operation.

%The output layer is designed differently according to the targets of the SB-T mapping and the SB-DFT mapping. For the SB-T mapping, a Linear full-connected (FC) layer is applied to transform the $\mathbf{E}^{e}$ to an $L$-dimensional tensor $\boldsymbol{w}^{'}(k)$:
%\begin{equation}
%\boldsymbol{w}^{'}(k) = \boldsymbol{G}_{time}\mathbf{E}^{e}+\boldsymbol{b}_{time},
%\end{equation} 
%where $\boldsymbol{G}_{time}$ and $\boldsymbol{b}_{time}$ are the weight and bias of the linear layer, respectively. Finally, $\widehat{\boldsymbol{w}}(k)$ is obtained by weighting $\boldsymbol{w}^{'}(k)$ with a Hanning window. 

%For SB-DFT mapping, 
After that, two FC layers were utilized to estimate the real and imaginary parts of the corresponding frequency response of $\widehat{\boldsymbol{W}}(k)$, i.e.,
\begin{equation}
\widehat{\boldsymbol{W}}^{r/i}_{DFT}(k) = \boldsymbol{G}_{DFT}^{r/i}\mathbf{E}^{e}+\boldsymbol{b}_{DFT}^{r/i},
\end{equation}
where  $\boldsymbol{G}_{DFT}^{r/i}$ and $\boldsymbol{b}_{DFT}^{r/i}$ are the weight and bias of the FC layer, respectively. The superscripts $r$ and $i$ denote the real and imaginary parts, respectively. $\widehat{\boldsymbol{w}}(k)$ could be \textcolor{black}{obtained} by calculating the inverse DFT (iDFT) of $\widehat{\boldsymbol{W}}_{DFT}(k)=\widehat{\boldsymbol{W}}^{r}_{DFT}(k) + j \widehat{\boldsymbol{W}}^{i}_{DFT}(k)$.

\subsection{\label{subsec:3:4}Signal Reconstruction}
To train the system in an end-to-end manner in the second stage, we \textcolor{black}{performed} the filtering operation in the frequency domain according to the convolution theorem. Specifically, we first segmented the noisy waveform into chunks of length $2P$ and hop size $r$, $\boldsymbol{x}^{filter}\in\mathbb{R}^{T\times2P}$. For the $k$th frame, the filtering result $\widehat{\boldsymbol{S}}(k)$ could be obtained by multiplying the DFT of $\boldsymbol{x}^{filter}(k)$ and $\widehat{\boldsymbol{W}}_{DFT}(k)$, given by
\begin{equation}
	\widehat{\boldsymbol{S}}(k) = DFT\{\boldsymbol{x}^{filter}(k)\}\widehat{\boldsymbol{W}}_{DFT}(k),
\end{equation}
%where $\boldsymbol{W}_{DFT}(k)$ is yielded by concatenating $\widehat{\boldsymbol{w}}(k)$ and a zeros tensor $\mathbf{O}\in\mathbb{R}^{L\times1}$ and then performing DFT when SB-T mapping is applied. 

Finally, the last $r$ samples of the iDFT of $\widehat{\boldsymbol{S}}(k)$ \textcolor{black}{were} selected as the enhanced output of the $k$th frame, given as:  
\begin{equation}
	\widehat{\boldsymbol{s}}(k) = iDFT\{\widehat{\boldsymbol{S}}(k)\}[2P-r+1:2P],
\end{equation}

\subsection{\label{subsec:3:5}Loss Function}
The mean squared error (MSE) $\mathcal{L}^{RI+Mag}$ was used as the loss function, which \textcolor{black}{led to} higher scores on speech quality and intelligibility metrics, that was
\begin{equation}
\label{eqn:equa2}
\mathcal{L}^{RI+Mag} = \mathcal{L}^{RI} + \mathcal{L}^{mag},
\end{equation} 
where
\begin{equation}
\label{eqn:equa3}
\mathcal{L}^{RI} = \left\| \widehat{S}_r - S_{r} \right\|^{2}_{F} + \left\| \widehat{S}_i - S_{i} \right\|^{2}_{F},
\end{equation} 
\begin{equation}
\label{eqn:equa4}
\mathcal{L}^{mag} = \left\| |\widehat{S}_r+j\widehat{S}_i| - |S| \right\|^{2}_{F},
\end{equation}  
where $\widehat{S}_r$ and $\widehat{S}_i$ are the predicted real and imaginary parts of the clean speech, and $|\cdot|$ extracts magnitude. $S_r$ and $S_i$ are the real and imaginary components of \textcolor{black}{the} clean speech. 

In this paper, a two-stage training strategy was applied to train the network. First, we trained the NR-Net with the CRM-based signal approximation loss. $\widehat{S}_r$ and $\widehat{S}_i$ were defined as $\widehat{\boldsymbol{W}}_{r}\boldsymbol{X}_{r}$ and $\widehat{\boldsymbol{W}}_{i}\boldsymbol{X}_{i}$, respectively. 

Then the parameters of NR-Net were frozen and we only train FA-Net. The MSE loss calculated \textcolor{black}{from} the filtered spectrum was back-propagated through all of the modules (including the signal construction module and the neural networks). 
 
\section{\label{sec:4} Experimental setup}
\subsection{\label{subsec:4:1} Dataset}
We \textcolor{black}{conducted} the experiments on the WSJ0-SI84 corpus \citep{paul1992design}, which includes 7138 utterances by 83 speakers (42 males and 41 females). Of these speakers, we set aside 6 speakers as untrained speakers, and 5428 and 957 utterances by 77 remaining speakers were chosen for training and validation, respectively. The noise clips were provided by the DNS-Challenge \citep{reddy2020interspeech} and we randomly selected 20,000 recordings as the noise set, \textcolor{black}{with a total duration of about} 55 hours. The noisy signal was generated as follows: a noise vector was generated by randomly \textcolor{black}{cutting} from the noise dataset, and then mixed with a randomly selected clean utterance at a randomly selected SNR. The SNR was set to range from -5 dB to 0 dB \textcolor{black}{in 1-dB steps}. As a result, we generated 50,000 and 4,000 noisy-clean pairs for training and validation, respectively. The training set contained around 110 hours of mixture while the validation set had around 7 hours. 

During the model evaluation, two test sets were created for each noise. One was \textcolor{black}{based on noise mixed with} clean speech utterances from 6 trained speakers, and the other \textcolor{black}{was based on noise mixed with untterances from} 6 untrained speakers to investigate speaker generalization capability of the \textcolor{black}{method}. Both test sets consisted of 3 males and 3 females. Three noise types were chosen for model evaluation, namely white Gaussian noise, babble noise, and factory1 noise from the NOISEX-92 dataset~\citep{varga1993assessment}. This selection included stationary noise, impulsive, as well as speech-like noise types. Note that all these three types of noise were untrained in the training stage. Four SNRs were set, namely -5 dB, 0 dB, 5 dB, and 10 dB. \textcolor{black}{In total,} 150 mixtures were generated with $25 \times 6$ utterances of 6 speakers for each case. 

\subsection{\label{subsec:4:2} Parameter Setup}
All the utterances were sampled at 16 kHz. The analysis filterbank had $M=512$ band-pass filters with the length of a prototype filter $L=512$, and a downsampling rate $r=64$. The model was trained for 50 epochs using the Adam \textcolor{black}{\citep{kingma2014adam}} optimizer. The initialized learning rate (LR) was set to 0.001, and we halved the LR when the validation loss did not increase for \textcolor{black}{two} consecutive epochs. The batch size was set to 16 at the utterance level, and the maximum utterance length was set to 8 seconds.

\subsection{\label{subsec:4:3} Comparison Systems}
We compared the performance of the proposed system with the following speech enhancement algorithms.
\begin{enumerate}[1)]
%	\item Noisy: speech corrupted with additive noise.
	\item MMSE-MA-FBE: the FBE presented in Sec.\ref{sec:2}. The subband-domain response was computed by the minimum mean-square error log-spectral amplitude (MMSE-LSA) estimator \citep{ephraim1985speech}, and the GDFT was applied to transform the response to the time-domain filter coefficients. Then the moving-average filter was utilized to approximate the time-domain filter of the FBE. To \textcolor{black}{ensure} the signal delay to 4 ms, the length of the MA filter $P$ was set to 128. 
	\item CRN-MA-FBE: the FBE was constructed using the subband-domain response of the FBE estimated by the NR-Net, followed by filter approximation using the MA filter with its length $P=128$.
	\item DeepFBE-T: the deep learning-based FBE implemented \textcolor{black}{the} domain transform by GDFT and used the DNN to learn the filter approximation mapping from time-domain filter coefficients of the FBE to the corresponding low-\textcolor{black}{order} filter. The NR-Net was first applied to estimate the subband-domain response of the FBE and then the GDFT of the response yields the time-domain filter coefficients of the FBE. The FA-Net was used to predict the corresponding low-\textcolor{black}{order} time-domain weighting factors, which were utilized to filter the mixture. \textcolor{black}{A 256-point DFT was implemented} since filter coefficients were padded \textcolor{black}{with} zeros to keep the same length as the input signal $\boldsymbol{x}^{filter}(k) \in \mathbb{R}^{2P \times 1}$, with $P=128$, in the frequency-domain signal reconstruction stage. Considering the symmetry of real-value filter coefficients in frequency, $D$ was set to 129 points.
	\item DeepFBE: the deep learning-based FBE described in Sec.\ref{sec:3}. \textcolor{black}{In contrast to} the DeepFBE-T, the FA-Net directly built the mapping from the subband-domain response of the FBE to the frequency-domain response of the corresponding low-\textcolor{black}{order} approximate filter. 
\end{enumerate}

\section{\label{sec:5} Results}
\subsection{\label{subsec:5:1} Objective measurements}
Three commonly used objective measurements were chosen to evaluate the performance of the proposed deep FBE, \textcolor{black}{including} segmental noise attenuation (segNA) \citep{fingscheidt2008environment}, perceptual evaluation of speech quality (PESQ) \citep{rix2001perceptual}, and segmental SNR (segSNR) \citep{loizou2007speech}. \textcolor{black}{For completeness,} \textcolor{black}{three composite measures proposed in \citep{loizou2007speech} with reference to ITU-T P.835 standard including CSIG, CBAK and COVL were also selected as objective metrics to evaluate these models.}

When computing the segNA, the residual noise $\hat n(t)$ needed to be separated from $\hat s(t)$. This was not a trivial task, and thus only noise-only frames were chosen to compute the segNA, \textcolor{black}{because $\hat n(t) = \hat s(t)$ in this case, the segNA could be obtained as}
\begin{equation}
	{segNA} = 10\log_{10}\left({\frac{1}{\mathcal{N_{\mathbb{F}}}}\sum_{m\in\mathbb{F}}{\frac{\sum_{\mu=0}^{r-1}{n^{2}(mr + \mu)}}{\sum_{\mu=0}^{r-1}{\widehat{n}^{2}(mr + \mu)}}}}\right),
\end{equation} 
where $\mathcal{N_{\mathbb{F}}}$ denotes the total number of noise-only frames and $\mathbb{F}$ indicates the indices of noise-only frames, i.e., $m\in\mathbb{F}$. The higher segNA, the less residual noise remains, and the better is the noise reduction performance of the method.

\textcolor{black}{The} PESQ score, ranging from -0.5 to 4.5, \textcolor{black}{was obtained from} the clean speech $s(t)$ and the enhanced speech $\widehat{s}(t)$. The higher the PESQ score, the better speech perceptual quality is.

The output segmental SNR is defined as:
\begin{equation}
	{segSNR}=\frac{10}{\mathcal{N}_{\mathbb{A}}}
	\sum_{m\in\mathbb{A}} {
		\log_{10}
		{
			\left(
			\frac{\sum_{\mu=0}^{r-1} {s^{2}(mr+\mu)} }
			{\sum_{\mu=0}^{r-1} {(\widehat{s}(mr+\mu)-s(mr+\mu))^2} }
		\right)
		}
	}, 
\end{equation}
where $\mathbb{A}$ means all frames.

For composite measures that aim to computationally approximate the Mean Opinion Score (MOS), the CSIG score is the MOS prediction of \textcolor{black}{perceived} signal distortion \textcolor{black}{based only on} the speech signal, the CBAK score measures the intrusiveness of background noise, and the COVL score represents the overall effect of the algorithm. All of them range from 1 to 5, \textcolor{black}{where higher scores indicate better performance}.

\subsection{\label{subsec:5:2} Objective Metrics for Trained Speakers}
\begin{figure}[t] 
	\centering
	\includegraphics[width=7.7cm]{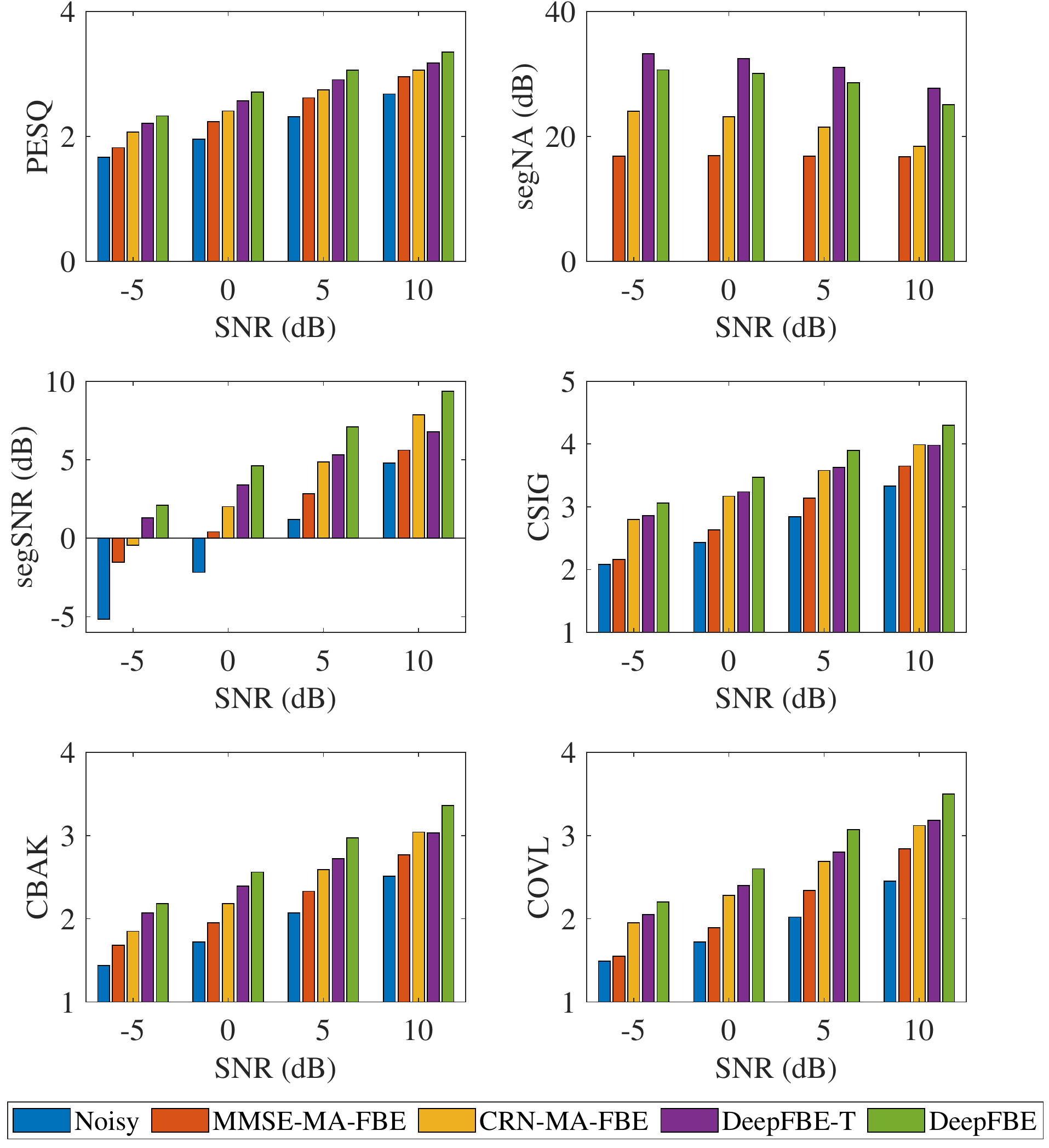}
	\caption{\label{fig:metrics_seen}{PESQ, segNA, segSNR, CSIG, CBAK and COVL scores under different SNRs for different methods in the trained speaker test set. Each value is averaged with three noise types.}}
\end{figure}

We evaluated and compared the systems using the WSJ0-SI84 trained speakers. Tables \ref{tbl:white}, \ref{tbl:babble}, and \ref{tbl:factory1} report the results in terms of PESQ, segNA, segSNR, CSIG, CBAK, and COVL for each case. \textcolor{black}{The average results across the three noises are shown} in Fig. \ref{fig:metrics_seen}.  
{
	\begin{table*}[t]
		\small
		\centering
		\caption{\label{tab:white} \textcolor{black}{Objective} result comparisons among different models in terms of PESQ, segNA, segSNR, CSIG, CBAK and COVL for white Gaussian noise in the trained speaker test set. \textcolor{black}{\textbf{BOLD} font indicates the best
		score in each case.}}
		
		\tabcolsep16pt
		\renewcommand\arraystretch{0.5}
		\begin{tabular}{clccccc}
			\hline\hline
			\multirow{2}{*}{Metrics}&\multirow{2}{*}{Methods} & \multicolumn{5}{c}{SNR (in dB)} \\
			\cline{3-7}
			
			{}&{} & -5 & 0 &5 &10 &Avg.  \\
			
			\hline
			\multirow{5}{*}{PESQ}
			&Noisy&1.59&1.90&2.28&2.68&2.11\\
			&MMSE-MA-FBE&1.84	&2.36	&2.78	&3.09	&2.52\\
			&CRN-MA-FBE&2.21	&2.55	&2.86	&3.15	&2.69\\
			&DeepFBE-T&2.36	&2.68	&2.98	&3.26	&2.82\\
			&DeepFBE&\textbf{2.52}	&\textbf{2.86}	&\textbf{3.16}	&\textbf{3.43}	&\textbf{2.99}\\
			\hline
			\multirow{5}{*}{segNA}
			&MMSE-MA-FBE&18.57	&18.32	&18.61	&18.51	&18.5\\
			&CRN-MA-FBE&25.07	&24.25	&21.78	&18.40	&22.54\\
			&DeepFBE-T&\textbf{34.59}	&\textbf{33.68}	&\textbf{31.94}	&\textbf{29.23}	&\textbf{32.36}\\
			&DeepFBE&31.97	&31.61	&30.18	&25.52	&29.82\\
			\hline
			\multirow{5}{*}{segSNR}
			&Noisy&-5.45	&-2.49	&0.86	&4.48	&-0.65\\
			&MMSE-MA-FBE&-1.43	&0.44	&2.99	&5.87	&1.97\\
			&CRN-MA-FBE&0.23	&2.78	&5.64	&8.67	&4.33\\
			&DeepFBE-T&2.18&	4.24	&6.03	&7.47	&4.98\\
			&DeepFBE&\textbf{2.74}	&\textbf{5.20}	&\textbf{7.47}	&\textbf{9.65}	&\textbf{6.27}\\
			\hline
			\multirow{5}{*}{CSIG}
			&Noisy&1.53	&1.94	&2.40	&2.94	&2.20\\
			&MMSE-MA-FBE&1.85	&2.47	&3.10	&3.65	&2.77\\
			&CRN-MA-FBE&2.84	&3.25	&3.65	&4.07	&3.45\\
			&DeepFBE-T&2.81	&3.20	&3.59	&3.95	&3.39\\
			&DeepFBE&\textbf{3.09}	&\textbf{3.52}	&\textbf{3.89}	&\textbf{4.23}	&\textbf{3.68}\\
			\hline
			\multirow{5}{*}{CBAK}
			&Noisy&1.50	&1.75	&2.05	&2.44	&1.93\\
			&MMSE-MA-FBE &1.77	&2.03	&2.44	&2.89	&2.28\\
			&CRN-MA-FBE&2.00	&2.36	&2.77	&3.22	&2.59\\
			&DeepFBE-T &2.18	&2.49	&2.81	&3.11	&2.65\\
			&DeepFBE &\textbf{2.31}	&\textbf{2.70}	&\textbf{3.07}	&\textbf{3.42}	&\textbf{2.87}\\
			\hline
			\multirow{5}{*}{COVL}
			&Noisy&1.24	&1.47	&1.76	&2.18	&1.66\\
			&MMSE-MA-FBE &1.42	&1.84	&2.39	&2.92	&2.14\\
			&CRN-MA-FBE &2.04	&2.41	&2.83	&3.28	&2.64\\
			&DeepFBE-T &2.07	&2.43	&2.82	&3.20	&2.63\\
			&DeepFBE &\textbf{2.29}	&\textbf{2.72}	&\textbf{3.13}	&\textbf{3.51}	&\textbf{2.91}\\
			\hline
			\hline
		\end{tabular}
		\label{tbl:white}
	\end{table*}
}

{
	\begin{table*}[t]
		\small
		\centering
		\caption{\label{tab:babble} \textcolor{black}{Objective} result comparisons among different models in terms of PESQ, segNA, segSNR, CSIG, CBAK and COVL for babble noise in the trained speaker test set.\textcolor{black}{\textbf{BOLD} font indicates the best
				score in each case.}}
		
		\tabcolsep16pt
		\renewcommand\arraystretch{0.5}
		\begin{tabular}{clccccc}
			\hline\hline
			\multirow{2}{*}{Metrics}&\multirow{2}{*}{Methods} & \multicolumn{5}{c}{SNR (in dB)} \\
			\cline{3-7}
			
			{}&{} & -5 & 0 &5 &10 &Avg.  \\
			
			\hline
			\multirow{5}{*}{PESQ}
			&Noisy&1.77&2.04&2.36&2.69&2.22\\
			&MMSE-MA-FBE&1.78&2.14&2.51&2.87&2.32\\
			&CRN-MA-FBE&1.96&2.30&2.66&2.97&2.47\\
			&DeepFBE-T&2.08&2.47&2.85&3.13&2.63\\
			&DeepFBE&\textbf{2.19}&\textbf{2.58}&\textbf{3.00}&\textbf{3.30}&\textbf{2.77}\\
			\hline
			\multirow{5}{*}{segNA}
			&MMSE-MA-FBE&15.52&15.55&15.57&15.33&15.49\\
			&CRN-MA-FBE&22.04&21.69&21.04&17.93&20.67\\
			&DeepFBE-T&\textbf{32.98}&\textbf{32.30}&\textbf{31.57}&\textbf{27.23}&\textbf{31.02}\\
			&DeepFBE&30.05&29.52&28.62&26.10&28.57\\
			\hline
			\multirow{5}{*}{segSNR}
			&Noisy&-4.99&-2.02&1.24&4.86&-0.23\\
			&MMSE-MA-FBE&-1.42&0.42&2.77&5.53&1.83\\
			&CRN-MA-FBE&-0.90&1.42&4.31&7.34&3.04\\
			&DeepFBE-T&0.54&2.67&4.71&6.25&3.54\\
			&DeepFBE&\textbf{1.60}&\textbf{4.14}&\textbf{6.80}&\textbf{9.17}&\textbf{5.43}\\
			\hline
			\multirow{5}{*}{CSIG}
			&Noisy&2.40&2.71&3.09&3.53&2.93\\
			&MMSE-MA-FBE&2.35&2.73&3.20&3.72&3.00\\
			&CRN-MA-FBE&2.75&3.12&3.53&3.90&3.33\\
			&DeepFBE-T&2.88&3.27&3.67&4.00&3.46\\
			&DeepFBE&\textbf{3.04}&\textbf{3.47}&\textbf{3.95}&\textbf{4.37}&\textbf{3.71}\\
			\hline
			\multirow{5}{*}{CBAK}
			&Noisy&1.42&1.71&2.07&2.52&1.93\\
			&MMSE-MA-FBE&1.63&1.90&2.26&2.72&2.13\\
			&CRN-MA-FBE&1.75&2.05&2.46&2.88&2.29\\
			&DeepFBE-T&1.98&2.30&2.66&2.96&2.48\\
			&DeepFBE&\textbf{2.09}&\textbf{2.47}&\textbf{2.92}&\textbf{3.34}&\textbf{2.70}\\
			\hline
			\multirow{5}{*}{COVL}
			&Noisy&1.65&1.86&2.16&2.58&2.06\\
			&MMSE-MA-FBE&1.63&1.92&2.34&2.85&2.19\\
			&CRN-MA-FBE&1.89&2.18&2.58&2.97&2.41\\
			&DeepFBE-T&2.04&2.39&2.80&3.16&2.60\\
			&DeepFBE&\textbf{2.15}&\textbf{2.55}&\textbf{3.06}&\textbf{3.52}&\textbf{2.82}\\
			\hline
			\hline
		\end{tabular}
		\label{tbl:babble}
	\end{table*}
}

{
	\begin{table*}[t]
		\small
		\centering
		\caption{\label{tab:factory1} \textcolor{black}{Objective} result comparisons among different methods in terms of PESQ, segNA, segSNR, CSIG, CBAK and COVL for factory1 noise in the trained speaker test set. \textcolor{black}{\textbf{BOLD} font indicates the best
				score in each case.}}
		
		\tabcolsep16pt
		\renewcommand\arraystretch{0.5}
		\begin{tabular}{clccccc}
			\hline\hline
			\multirow{2}{*}{Metrics}&\multirow{2}{*}{Methods} & \multicolumn{5}{c}{SNR (in dB)} \\
			\cline{3-7}
			
			{}&{} & -5 & 0 &5 &10 &Avg.  \\
			
			\hline
			\multirow{5}{*}{PESQ}
			&Noisy &1.64 &1.95 &2.33 &2.68 &2.15\\
			&MMSE-MA-FBE &1.84 &2.21 &2.59 &2.92 &2.39\\
			&CRN-MA-FBE &2.03 &2.39 &2.74 &3.04 &2.55\\
			&DeepFBE-T &2.18 &2.55 &2.89 &3.15 &2.69\\
			&DeepFBE &\textbf{2.26} &\textbf{2.67} &\textbf{3.03} &\textbf{3.31} &\textbf{2.82}\\
			\hline
			\multirow{5}{*}{segNA}
			&MMSE-MA-FBE &16.51 &16.98 &16.41 &16.53 &16.61\\
			&CRN-MA-FBE &24.46 &23.57 &21.64 &18.86 &22.13\\
			&DeepFBE-T &\textbf{32.14} &\textbf{31.45} &\textbf{29.73} &\textbf{26.73} &\textbf{30.01}\\
			&DeepFBE &30.03 &29.15 &27.11 &23.62 &27.48\\
			\hline
			\multirow{5}{*}{segSNR}
			&Noisy &-5.01 &-2.04 &1.45 &4.99 &-0.15\\
			&MMSE-MA-FBE &-1.71 &0.35 &2.72 &5.43 &1.70\\
			&CRN-MA-FBE &-0.67 &1.82 &4.68 &7.59 &3.35\\
			&DeepFBE-T &1.19 &3.30 &5.15 &6.62 &4.06\\
			&DeepFBE &\textbf{1.97} &\textbf{4.53} &\textbf{7.00} &\textbf{9.28} &\textbf{5.69}\\
			\hline
			\multirow{5}{*}{CSIG}
			&Noisy &2.31 &2.64 &3.04 &3.52 &2.88\\
			&MMSE-MA-FBE &2.29 &2.68 &3.12 &3.59 &2.92\\
			&CRN-MA-FBE &2.79 &3.16 &3.57 &4.00 &3.38\\
			&DeepFBE-T &2.89 &3.25 &3.63 &3.99 &3.44\\
			&DeepFBE &\textbf{3.04} &\textbf{3.43} &\textbf{3.87} &\textbf{4.29} &\textbf{3.66}\\
			\hline
			\multirow{5}{*}{CBAK}
			&Noisy &1.42 &1.71 &2.09 &2.56 &1.94\\
			&MMSE-MA-FBE &1.65 &1.92 &2.27 &2.71 &2.14\\
			&CRN-MA-FBE &1.79 &2.13 &2.55 &3.01 &2.37\\
			&DeepFBE-T &2.05 &2.36 &2.69 &3.01 &2.53\\
			&DeepFBE &\textbf{2.15} &\textbf{2.51} &\textbf{2.93} &\textbf{3.33} &\textbf{2.73}\\
			\hline
			\multirow{5}{*}{COVL}
			&Noisy &1.59 &1.81 &2.13 &2.58 &2.03\\
			&MMSE-MA-FBE &1.61 &1.90 &2.28 &2.76 &2.14\\
			&CRN-MA-FBE &1.92 &2.24 &2.66 &3.12 &2.49\\
			&DeepFBE-T &2.06 &2.39 &2.78 &3.18 &2.60\\
			&DeepFBE &\textbf{2.17} &\textbf{2.54} &\textbf{3.01} &\textbf{3.47} &\textbf{2.80}\\
			\hline
			\hline
		\end{tabular}
		\label{tbl:factory1}
	\end{table*}
}

Several observations can be made. First, compared with MMSE-MA-FBE, the CRN-MA-FBE achieved better performance, \textcolor{black}{especially for low SNRs} and non-stationary noise scenarios. This is because it is difficult for the conventional noise estimation method to track rapid changes of noise power, resulting in speech distortion and remaining residual noise. For example, for babble noise, \textcolor{black}{the CSIG scores of the MMSE-MA-FBE were lower than for} the noisy mixture at SNR=-5dB, \textcolor{black}{for which} the MMSE-MA-FBE approach did not work at all, getting PESQ=1.78 and COVL=1.63, which was almost the same as the PESQ value of 1.77 and the COVL value of 1.63 for noisy speech. The CRN could suppress non-stationary noise better, and also achieved lower speech distortion. Besides, when a DNN replaced the MA filter approximation, it showed consistent improvements for all metrics over the competing methods for babble noise and factory1 noise, indicating that the filter approximated by the data-adaptive mapping was more beneficial for the speech enhancement task than a fixed filter design method. Finally, compared with DeepFBE-T, DeepFBE consistently outperformed DeepFBE-T in terms of PESQ, segSNR, CSIG, CBAK, and COVL. For speech distortion, compared with DeepFBE-T, DeepFBE achieved \textcolor{black}{a 0.25 CSIG score} improvement on average. For noise reduction, around 0.21 average improvement in CBAK is \textcolor{black}{obtained, although the segNA decreased 2.51 dB on average}. The reason is that segNA was computed with noise-only segments and DeepFBE-T had a powerful noise suppression capability for non-speech frames. Average 0.15, 1.60 dB and 0.23 increase in PESQ, segSNR and COVL scores were achieved, which showed that the quality of enhanced speech with DeepFBE could be improved. This indicated that a non-constrained adaptive filter design was able to approximate the desired filter response better than a constrained method.
%, which is also found in acoustic echo cancellation (AEC) and active noise control (ANC) areas.   

\subsection{\label{subsec:5:3} Objective Metrics for Untrained Speakers}
\begin{figure}[t] 
	\centering
	\includegraphics[width=7.7cm]{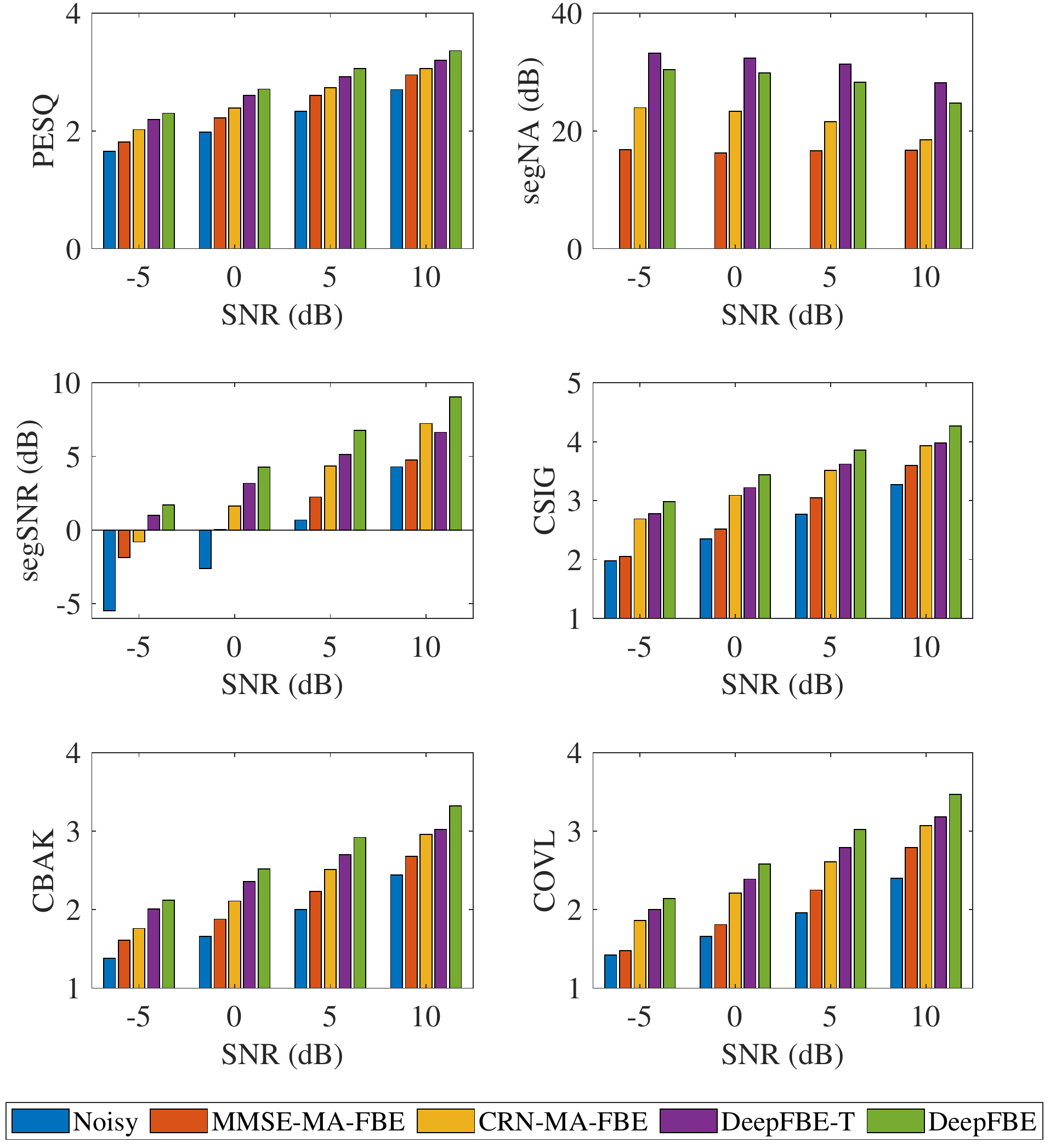}
	\caption{\label{fig:metrics_unseen}{PESQ, segNA, segSNR, CSIG, CBAK and COVL scores under different SNRs for different methods in the untrained speaker test set. Each value is averaged with three noise types.}}
\end{figure}

The evaluation results \textcolor{black}{for untrained speakers are illustrated in Fig. \ref{fig:metrics_unseen}, based on values averaged across the} three types of noise. The results were consistent with \textcolor{black}{those for} the trained condition, showing that the proposed DeepFBE algorithm generalized very well \textcolor{black}{to unseen speakers}. Because less speech distortion and more noise reduction could be obtained by applying DNN to the FBE system, our system performed better than the competing methods, as expected. For example, the average difference in performance between DeepFBE and CRN-MA-FBE was 6.49 dB segNA, 2.36 dB segSNR, 0.31 PESQ, 0.33 CSIG, 0.39 CBAK and 0.36 COVL. These results indicated that utilizing the deep learning-based method to replace the traditional FBE results in better speech denoising performance with the same signal latency, which confirmed the effectiveness and generalization of the proposed deep-learning-based low-latency speech enhancement framework.

\subsection{\label{subsec:5:4} Spectrogram Analysis}
\begin{figure}[t] 
	\centering
	\includegraphics[width=7.7cm]{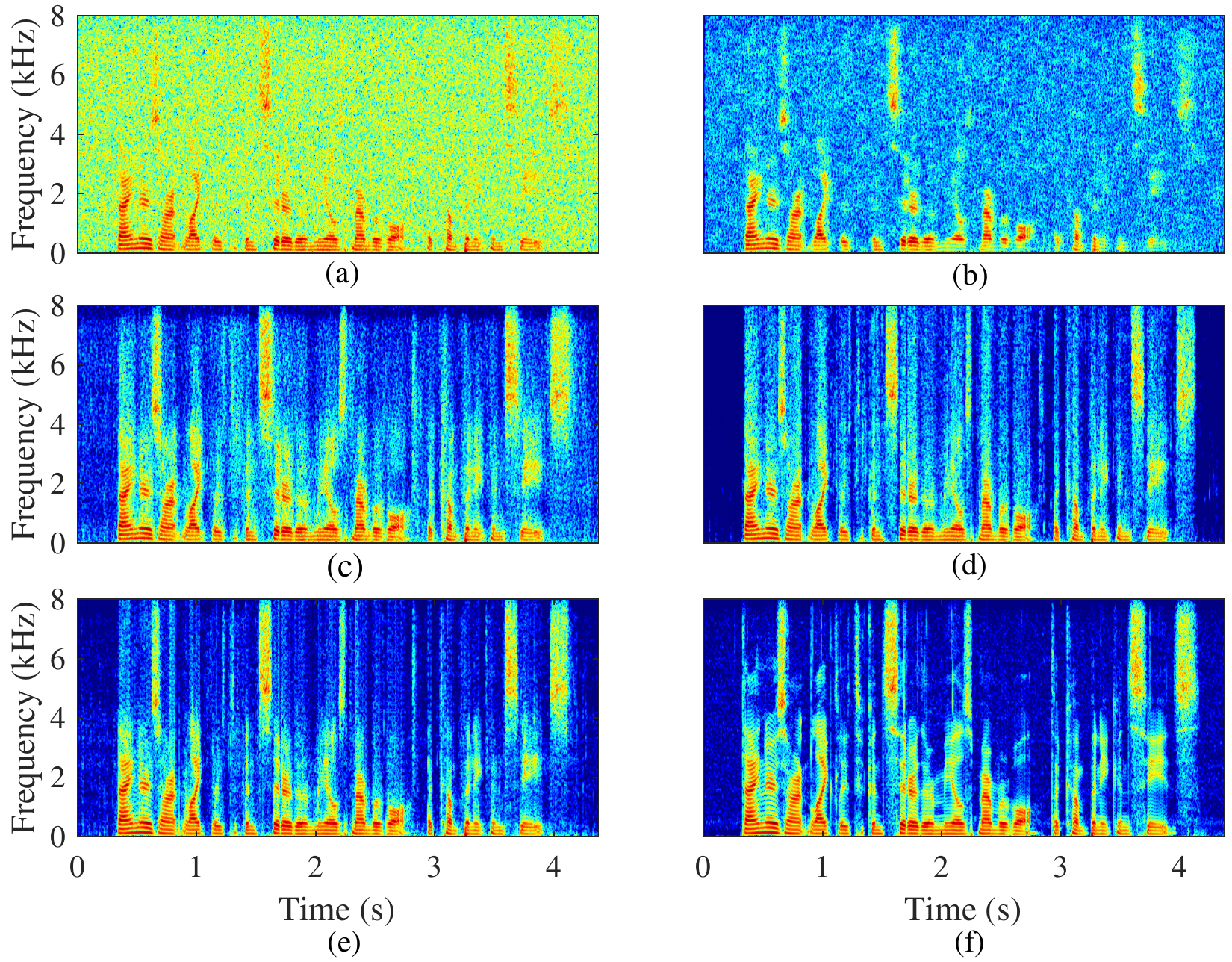}
	\caption{\label{fig:spec_white}{Spectrograms of (a) noisy speech corrupted with white Gaussian noise \textcolor{black}{at 0-dB SNR}, and enhanced speech processed by (b) MMSE-MA-FBE, (c) CRN-MA-FBE, (d) DeepFBE-T and (e) DeepFBE, (f) clean speech.}}
\end{figure}
\begin{figure}[t] 
	\centering
	\includegraphics[width=7.7cm]{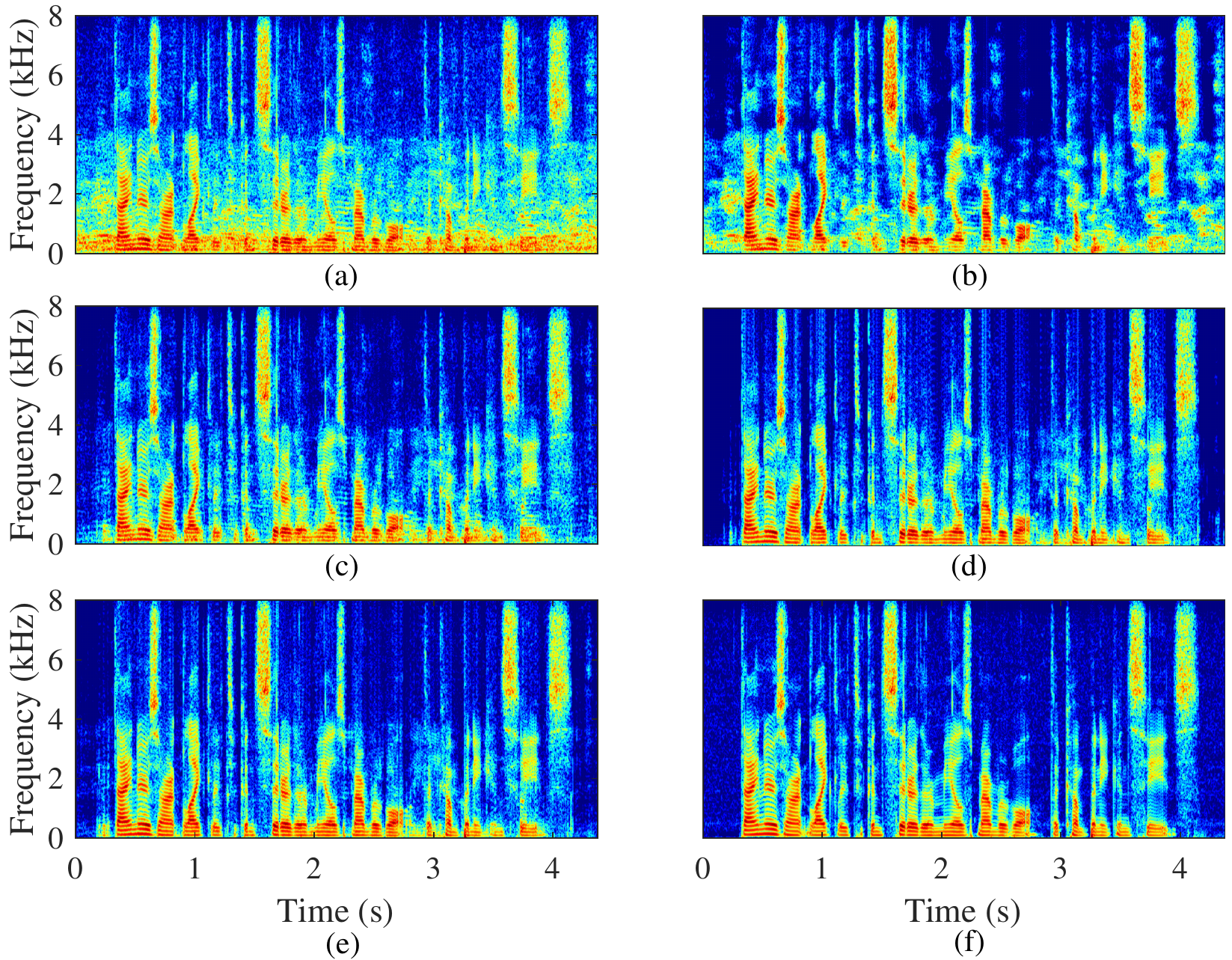}
	\caption{\label{fig:spec_babble}{Spectrograms of (a) noisy speech corrupted with babble noise \textcolor{black}{at 0-dB SNR}, and enhanced speech processed by (b) MMSE-MA-FBE, (c) CRN-MA-FBE, (d) DeepFBE-T and (e) DeepFBE, (f) clean speech.}}
\end{figure}
\begin{figure}[t] 
	\centering
	\includegraphics[width=7.7cm]{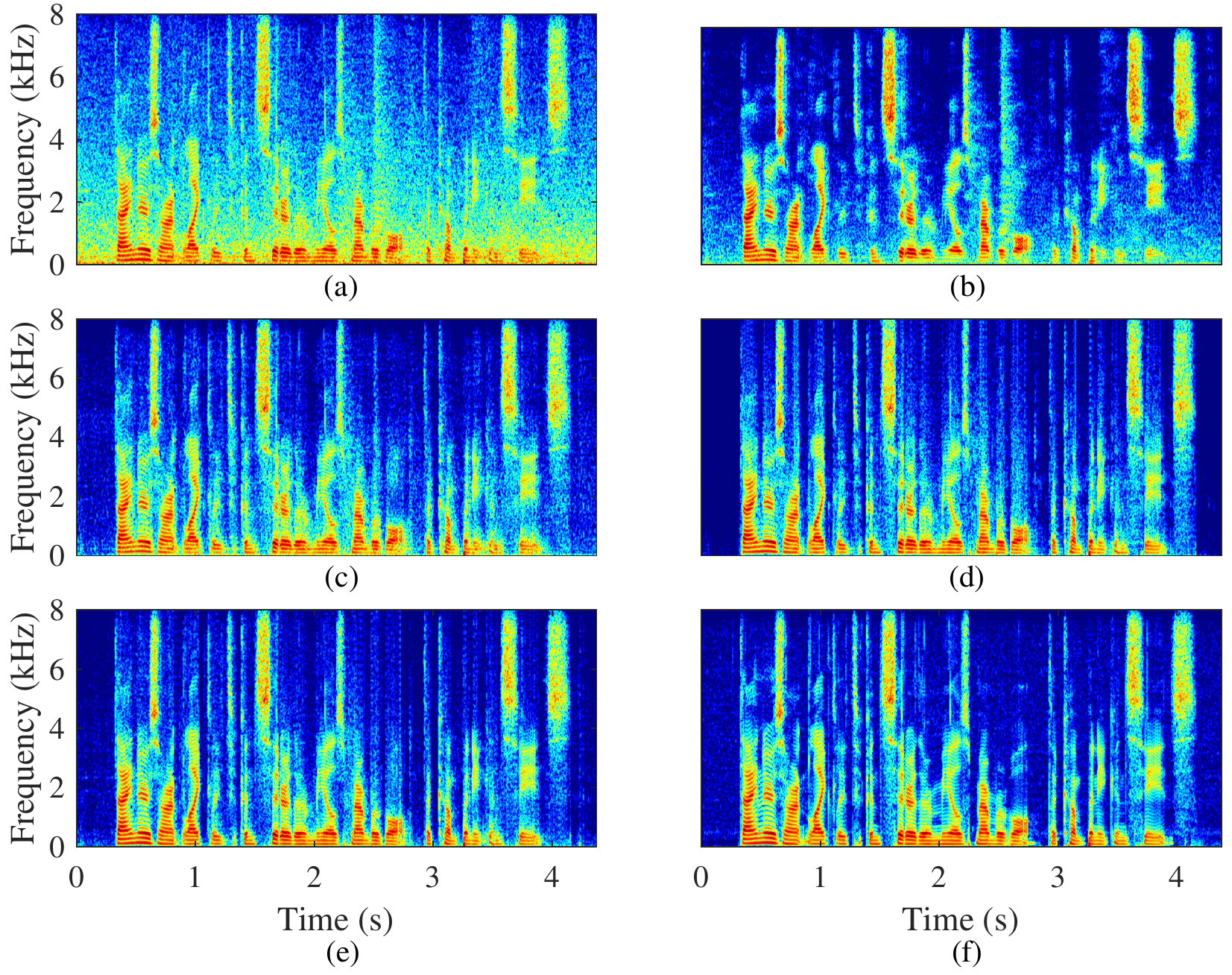}
	\caption{\label{fig:spec_factory1}{Spectrograms of (a) noisy speech corrupted with factory1 noise \textcolor{black}{at 0-dB SNR}, and enhanced speech processed by (b) MMSE-MA-FBE, (c) CRN-MA-FBE, (d) DeepFBE-T and (e) DeepFBE, (f) clean speech.}}
\end{figure}
Finally, we analyzed the enhanced speech spectrograms processed by different methods. Fig. \ref{fig:spec_white} (a) shows the spectrogram of the speech corrupted by white Gaussian noise \textcolor{black}{at 0-dB SNR. Performance} was improved when we replaced the MMSE-LSA with the CRN, which showed that the DNN could model the complex non-linear relationship from the subband-domain \textcolor{black}{features of noisy speech to the subband-domain features} of clean speech. Compared with the benchmarks, DeepFBE-T could better suppress noise \textcolor{black}{during} noise-only segments, which was consistent with the test results of the segNA metric, and DeepFBE showed the best performance overall. In non-stationary noise cases (see Figure \ref{fig:spec_babble} (b) and \ref{fig:spec_factory1} (b) for details), MMSE-MA-FBE preserved a lot of residual background noise components and had significant speech distortion, which \textcolor{black}{happened} because the noise estimation of the MMSE-LSA method could not track \textcolor{black}{large} changes of non-stationary noise power spectral density, while the proposed method performed well in non-stationary noise scenarios.

\section{\label{sec:6} Conclusion}
In this paper, we proposed a deep learning-based filter-bank equalizer namely DeepFBE for low-latency speech enhancement. A subband adaptive filtering technique was applied to reduce the signal latency while maintaining the high frequency resolution. First, the subband-domain response was estimated by the NR-Net to suppress background noise. Then the FA-Net was utilized to build the mapping between the response \textcolor{black}{in the} subband-domain and the frequency-domain response of a low-\textcolor{black}{order} filter. We conducted experiments \textcolor{black}{using} the WSJ0-SI84 dataset. The results demonstrated that the proposed framework significantly outperformed the benchmarks in both stationary and non-stationary noise situations, and had good generalization toward untrained speakers. In the future, we aim to explore the \textcolor{black}{work on model compression for DeepFBE} to satisfy the low power consumption requirement of hearing aids, and investigate the effectiveness of our system in noisy and \textcolor{black}{reverberant} scenarios.
%% before appendix (optional) and bibliography:
% -------------------------------------------------------------------------------------------------------------------
%   Appendix  (optional)

%\appendix
%\section{Appendix title}

%If only one appendix, please use
%\appendix*
%\section{Appendix title}
%=======================================================
%Use \bibliography{<name of your .bib file>}+

\begin{thebibliography}{22}
\def\enquote#1{``#1,''}
\def\plainquote#1{``#1''}
\expandafter\ifx\csname natexlab\endcsname\relax\def\natexlab#1{#1}\fi
\providecommand{\dourl}[1]{\href{http://#1}{\nolinkurl{#1}}}
\providecommand{\bibinfo}[2]{#2}
\providecommand{\noopsort}[1]{}
\providecommand{\switchargs}[2]{#2#1}
  \def\eatspace #1{#1}
\bibitem[{Alc{\'a}ntara \emph{et~al.}(2003)}]{alcantara2003evaluation}
\textcolor{black}{\bibinfo{author}{Alc{\'a}ntara, J. I.}, \bibinfo{author}{ Moore, B. C. J.},
\bibinfo{author}{K{\"u}hnel, V.}, and \bibinfo{author}{Launer, S.} (\textbf{\bibinfo{year}{2003}}).
  \enquote{\bibinfo{title}{Evaluation of the noise reduction system in a commercial digital hearing aid}}
  \bibinfo{journal}{International Journal of Audiology} \textbf{42}(1), \bibinfo{pages}{34--42}.}

\bibitem[{Ephraim and Malah(1985)}]{ephraim1985speech}
\bibinfo{author}{Ephraim, Y.},  and \bibinfo{author}{Malah, D.}
  (\textbf{\bibinfo{year}{1985}}). \enquote{\bibinfo{title}{Speech enhancement
  using a minimum mean-square error log-spectral amplitude estimator}}
  \bibinfo{journal}{IEEE Transactions on Acoustics, Speech, and Signal
  Processing} \textbf{33}(2), \bibinfo{pages}{443--445}.

\bibitem[{Fingscheidt \emph{et~al.}(2008)Fingscheidt, Suhadi, and
  Stan}]{fingscheidt2008environment}
\bibinfo{author}{Fingscheidt, T.}, \bibinfo{author}{Suhadi, S.},  and
  \bibinfo{author}{Stan, S.} (\textbf{\bibinfo{year}{2008}}).
  \enquote{\bibinfo{title}{Environment-optimized speech enhancement}}
  \bibinfo{journal}{IEEE Transactions on Audio, Speech, and Language
  Processing} \textbf{16}(4), \bibinfo{pages}{825--834}.

\bibitem[{Holube \emph{et~al.}(1999)Holube, I and Hamacher, V and Wesselkamp, M}]{Holube1999}
\textcolor{black}{\bibinfo{author}{Holube, I.}, \bibinfo{author}{Hamacher, V.}, and
  \bibinfo{author}{Wesselkamp, M.} (\textbf{\bibinfo{year}{1999}}).
  \enquote{\bibinfo{title}{Hearing Instruments: noise reduction strategies}} \bibinfo{journal}{in Proc. 18th Danavox Symposium: Auditory Models and Non-linear Hearing Instruments} \bibinfo{pages}{359--377}.}


\bibitem[{Hu \emph{et~al.}(2020)Hu, Liu, Lv, Xing, Zhang, Fu, Wu, Zhang, and
  Xie}]{hu2020dccrn}
\bibinfo{author}{Hu, Y.}, \bibinfo{author}{Liu, Y.}, \bibinfo{author}{Lv, S.},
  \bibinfo{author}{Xing, M.}, \bibinfo{author}{Zhang, S.}, \bibinfo{author}{Fu,
  Y.}, \bibinfo{author}{Wu, J.}, \bibinfo{author}{Zhang, B.},  and
  \bibinfo{author}{Xie, L.} (\textbf{\bibinfo{year}{2020}}).
  \enquote{\bibinfo{title}{DCCRN: Deep complex convolution recurrent network
  for phase-aware speech enhancement}} \bibinfo{journal}{in Proc. Interspeech
  2020} \bibinfo{pages}{2472--2476}.

\bibitem[{Hummersone \emph{et~al.}(2014)Hummersone, Stokes, and
  Brookes}]{Hummersone2014}
\bibinfo{author}{Hummersone, C.}, \bibinfo{author}{Stokes, T.},  and
  \bibinfo{author}{Brookes, T.} (\textbf{\bibinfo{year}{2014}}).
  \emph{\bibinfo{title}{On the Ideal Ratio Mask as the Goal of Computational
  Auditory Scene Analysis}}, \bibinfo{pages}{349--368}
  (\bibinfo{publisher}{Springer Berlin Heidelberg}, \bibinfo{address}{Berlin,
  Heidelberg}), \dourl{https://doi.org/10.1007/978-3-642-55016-4_12},
  \dodoi{10.1007/978-3-642-55016-4_12}.

\bibitem[{Kates and Arehart(2005)}]{kates2005multichannel}
\bibinfo{author}{Kates, J.~M.},  and \bibinfo{author}{Arehart, K.~H.}
  (\textbf{\bibinfo{year}{2005}}). \enquote{\bibinfo{title}{Multichannel
  dynamic-range compression using digital frequency warping}}
  \bibinfo{journal}{EURASIP Journal on Advances in Signal Processing}
  \textbf{2005}(18), \bibinfo{pages}{1--12}.

\bibitem[{Kingma \emph{et~al.}(2015)Kingma, and Ba}]{kingma2014adam}
\textcolor{black}{\bibinfo{author}{Kingma, D.~P.}, and \bibinfo{author}{Ba, L.~J.} 
(\textbf{\bibinfo{year}{2015}}). \enquote{\bibinfo{title}{Adam: A method for stochastic optimization}} \bibinfo{journal}{in {\it International Conference on Learning Representations} ({\it ICLR}), 2015}.}

\bibitem[{Loizou(2007)}]{loizou2007speech}
\bibinfo{author}{Loizou, P.~C.} (\textbf{\bibinfo{year}{2007}}).
  \emph{\bibinfo{title}{Speech enhancement: theory and practice}}
  (\bibinfo{publisher}{CRC press}).

\bibitem[{L{\"o}llmann and Vary(2007)}]{lollmann2007uniform}
\bibinfo{author}{L{\"o}llmann, H.~W.},  and \bibinfo{author}{Vary, P.}
  (\textbf{\bibinfo{year}{2007}}). \enquote{\bibinfo{title}{Uniform and warped
  low delay filter-banks for speech enhancement}} \bibinfo{journal}{Speech
  Communication} \textbf{49}(7-8), \bibinfo{pages}{574--587}.

\bibitem[{Nakaoka \emph{et~al.}(2021)Nakaoka, Li, Inoue, and
  Makino}]{nakaoka2021teacher}
\bibinfo{author}{Nakaoka, S.}, \bibinfo{author}{Li, L.},
  \bibinfo{author}{Inoue, S.},  and \bibinfo{author}{Makino, S.}
  (\textbf{\bibinfo{year}{2021}}). \enquote{\bibinfo{title}{Teacher-student
  learning for low-latency online speech enhancement using wave-u-net}} in
  \emph{\bibinfo{booktitle}{ICASSP 2021-2021 IEEE International Conference on
  Acoustics, Speech and Signal Processing (ICASSP)}},
  \bibinfo{organization}{IEEE}, pp. \bibinfo{pages}{661--665}.

\bibitem[{Paliwal \emph{et~al.}(2011)Paliwal, W{\'o}jcicki, and
  Shannon}]{paliwal2011importance}
\bibinfo{author}{Paliwal, K.}, \bibinfo{author}{W{\'o}jcicki, K.},  and
  \bibinfo{author}{Shannon, B.} (\textbf{\bibinfo{year}{2011}}).
  \enquote{\bibinfo{title}{The importance of phase in speech enhancement}}
  \bibinfo{journal}{Speech Communication} \textbf{53}(4),
  \bibinfo{pages}{465--494}.

\bibitem[{Paul and Baker(1992)}]{paul1992design}
\bibinfo{author}{Paul, D.~B.},  and \bibinfo{author}{Baker, J.}
  (\textbf{\bibinfo{year}{1992}}). \enquote{\bibinfo{title}{The design for the
  wall street journal-based csr corpus}} in \emph{\bibinfo{booktitle}{Speech
  and Natural Language: Proceedings of a Workshop Held at Harriman, New York,
  February 23-26, 1992}}.

\bibitem[{Popelka \emph{et~al.}(2016)Popelka, Moore, Fay, and
  Popper}]{popelka2016hearing}
\bibinfo{author}{Popelka, G.~R.}, \bibinfo{author}{Moore, B.~C.},
  \bibinfo{author}{Fay, R.~R.},  and \bibinfo{author}{Popper, A.~N.}
  (\textbf{\bibinfo{year}{2016}}). \emph{\bibinfo{title}{Hearing aids}}
  (\bibinfo{publisher}{Springer}).

\bibitem[{Proakis and Manolakis(1996)}]{proakis1996digital}
\bibinfo{author}{Proakis, J.~G.},  and \bibinfo{author}{Manolakis, D.~G.}
  (\textbf{\bibinfo{year}{1996}}). \emph{\bibinfo{title}{Digital signal
  processing: Principles, Algorithms, and Applications}}
  (\bibinfo{publisher}{Prentice Hall}).

\bibitem[{Reddy \emph{et~al.}(2020)Reddy, Beyrami, Dubey, Gopal, Cheng, Cutler,
  Matusevych, Aichner, Aazami, Braun \emph{et~al.}}]{reddy2020interspeech}
\bibinfo{author}{Reddy, C.~K.}, \bibinfo{author}{Beyrami, E.},
  \bibinfo{author}{Dubey, H.}, \bibinfo{author}{Gopal, V.},
  \bibinfo{author}{Cheng, R.}, \bibinfo{author}{Cutler, R.},
  \bibinfo{author}{Matusevych, S.}, \bibinfo{author}{Aichner, R.},
  \bibinfo{author}{Aazami, A.}, \bibinfo{author}{Braun, S.} \emph{et~al.}
  (\textbf{\bibinfo{year}{2020}}). \enquote{\bibinfo{title}{The interspeech
  2020 deep noise suppression challenge: Datasets, subjective speech quality
  and testing framework}} \bibinfo{journal}{arXiv preprint arXiv:2001.08662} .

\bibitem[{Rix \emph{et~al.}(2001)Rix, Beerends, Hollier, and
  Hekstra}]{rix2001perceptual}
\bibinfo{author}{Rix, A.~W.}, \bibinfo{author}{Beerends, J.~G.},
  \bibinfo{author}{Hollier, M.~P.},  and \bibinfo{author}{Hekstra, A.~P.}
  (\textbf{\bibinfo{year}{2001}}). \enquote{\bibinfo{title}{Perceptual
  evaluation of speech quality (PESQ)-a new method for speech quality
  assessment of telephone networks and codecs}} in
  \emph{\bibinfo{booktitle}{2001 IEEE International Conference on Acoustics,
  Speech, and Signal Processing. Proceedings (Cat. No. 01CH37221)}},
  \bibinfo{organization}{IEEE}, Vol. 2, pp. \bibinfo{pages}{749--752}.

\bibitem[{Roman \emph{et~al.}(2003)Roman, Wang, and
  Brown}]{Roman03speechsegregation}
\bibinfo{author}{Roman, N.}, \bibinfo{author}{Wang, D.},  and
  \bibinfo{author}{Brown, G.~J.} (\textbf{\bibinfo{year}{2003}}).
  \enquote{\bibinfo{title}{Speech segregation based on sound localization}}
  \bibinfo{journal}{The Journal of the Acoustical Society of America}
  \textbf{114}(4), \bibinfo{pages}{2236--2252}.

\bibitem[{Stone \emph{et~al.}(1999)Stone, and Moore}]{Stone99tolerable}
\textcolor{black}{\bibinfo{author}{Stone, M. A.}, and \bibinfo{author}{Moore, B. C. J.}, (\textbf{\bibinfo{year}{1999}}).
\enquote{\bibinfo{title}{Tolerable hearing-aid delays. I. Estimation of limits imposed by the auditory path alone using simulated hearing losses}}
\bibinfo{journal}{Ear Hear}
\textbf{20}, \bibinfo{pages}{182--192}.
}

\bibitem[{Stone \emph{et~al.}(2008)Stone, Moore, Meisenbacher, and Derleth}]{Stone08tolerable}
\textcolor{black}{\bibinfo{author}{Stone, M. A.}, \bibinfo{author}{Moore, B. C. J.}, \bibinfo{author}{Meisenbacher, K.}, and \bibinfo{author}{Derleth, R. P.}, (\textbf{\bibinfo{year}{2008}}).
\enquote{\bibinfo{title}{Tolerable hearing-aid delays. V.  Estimation of limits for open canal fittings}}
\bibinfo{journal}{Ear Hear}
\textbf{29}, \bibinfo{pages}{601--617}.}

\bibitem[{Tan \emph{et~al.}(2018)Tan, Chen, and Wang}]{tan2018gated}
\bibinfo{author}{Tan, K.}, \bibinfo{author}{Chen, J.},  and
  \bibinfo{author}{Wang, D.} (\textbf{\bibinfo{year}{2018}}).
  \enquote{\bibinfo{title}{Gated residual networks with dilated convolutions
  for monaural speech enhancement}} \bibinfo{journal}{IEEE/ACM Transactions on
  Audio, Speech, and Language Processing} \textbf{27}(1),
  \bibinfo{pages}{189--198}.

\bibitem[{Tan and Wang(2020)}]{tan2019learning}
\bibinfo{author}{Tan, K.},  and \bibinfo{author}{Wang, D.}
  (\textbf{\bibinfo{year}{2020}}). \enquote{\bibinfo{title}{Learning complex
  spectral mapping with gated convolutional recurrent networks for monaural
  speech enhancement}} \bibinfo{journal}{IEEE/ACM Transactions on Audio,
  Speech, and Language Processing} \textbf{28}, \bibinfo{pages}{380--390}.

\bibitem[{Varga and Steeneken(1993)}]{varga1993assessment}
\bibinfo{author}{Varga, A.},  and \bibinfo{author}{Steeneken, H.~J.}
  (\textbf{\bibinfo{year}{1993}}). \enquote{\bibinfo{title}{Assessment for
  automatic speech recognition: II. NOISEX-92: A database and an experiment to
  study the effect of additive noise on speech recognition systems}}
  \bibinfo{journal}{Speech Communication} \textbf{12}(3),
  \bibinfo{pages}{247--251}.

\bibitem[{Vary(2006)}]{petervary2006FBE}
\bibinfo{author}{Vary, P.} (\textbf{\bibinfo{year}{2006}}).
  \enquote{\bibinfo{title}{An adaptive filter-bank equalizer for speech
  enhancement}} \bibinfo{journal}{Signal Processing} \textbf{86}(6),
  \bibinfo{pages}{1206--1214}.

\bibitem[{Wang \emph{et~al.}(2021)Wang, Naithani, Politis, and
  Virtanen}]{wang2021deep}
\bibinfo{author}{Wang, S.}, \bibinfo{author}{Naithani, G.},
  \bibinfo{author}{Politis, A.},  and \bibinfo{author}{Virtanen, T.}
  (\textbf{\bibinfo{year}{2021}}). \enquote{\bibinfo{title}{Deep neural network
  based low-latency speech separation with asymmetric analysis-synthesis window
  pair}} \bibinfo{journal}{arXiv preprint arXiv:2106.11794} .

\bibitem[{Williamson \emph{et~al.}(2015)Williamson, Wang, and
  Wang}]{williamson2015complex}
\bibinfo{author}{Williamson, D.~S.}, \bibinfo{author}{Wang, Y.},  and
  \bibinfo{author}{Wang, D.} (\textbf{\bibinfo{year}{2015}}).
  \enquote{\bibinfo{title}{Complex ratio masking for monaural speech
  separation}} \bibinfo{journal}{IEEE/ACM Transactions on Audio, Speech, and
  Language Processing} \textbf{24}(3), \bibinfo{pages}{483--492}.

\bibitem[{Xu \emph{et~al.}(2014)Xu, Du, Dai, and Lee}]{xu2014regression}
\bibinfo{author}{Xu, Y.}, \bibinfo{author}{Du, J.}, \bibinfo{author}{Dai,
  L.-R.},  and \bibinfo{author}{Lee, C.-H.} (\textbf{\bibinfo{year}{2014}}).
  \enquote{\bibinfo{title}{A regression approach to speech enhancement based on
  deep neural networks}} \bibinfo{journal}{IEEE/ACM Transactions on Audio,
  Speech, and Language Processing} \textbf{23}(1), \bibinfo{pages}{7--19}.

\end{thebibliography}
%to make your bibliography with BibTeX. 
\section*{References}

\end{document}